\documentclass[twocolumn,prc,showpacs,showkeys,preprintnumbers,unsortedaddress,amsmath,amssymb,floatfix]{revtex4}
\input epsf.sty
\usepackage{graphicx}
\usepackage{bm}

\newcommand{\be}{\begin{equation}}
\newcommand{\ee}{\end{equation}}
\newcommand{\bea}{\begin{eqnarray}}
\newcommand{\eea}{\end{eqnarray}}

\begin{document}

\title{Microscopic Study of the Isoscalar Giant Monopole Resonance in Cd, Sn and Pb Isotopes}

\author{Li-Gang Cao$^{1,2,3}$, H. Sagawa$^{2,4}$, G. Col$\grave{\text{o}}$$^{5,6}$}

\affiliation{${}^1$Institute of Modern Physics, Chinese Academy of
Science, Lanzhou 730000, P.R. China}

\affiliation{${}^2$Center for Mathematics and Physics, University of Aizu,
Aizu-Wakamatsu, Fukushima 965-8560, Japan}

\affiliation{${}^3$Center of Theoretical Nuclear Physics,
National Laboratory of Heavy Ion Accelerator of Lanzhou, Lanzhou
730000, P.R. China}

\affiliation{${}^4$RIKEN, Nishina Center,
Wako, 351-0198 ,Japan}

\affiliation{${}^5$Dipartimento di Fisica, Universit$\grave{\text{a}}$
degli Studi di Milano, via Celoria 16, 20133 Milano, Italy}

\affiliation{${}^6$Istituto Nazionale di Fisica Nucleare (INFN), Sez. di
Milano, via Celoria 16, 20133 Milano, Italy}

\begin{abstract}
The isoscalar giant monopole resonance (ISGMR) in Cd, Sn and Pb
isotopes has been studied within the self-consistent Skyrme
Hartree-Fock+BCS and quasi-particle random phase approximation
(QRPA). Three Skyrme parameter sets are used in the
calculations, i.e., SLy5, SkM$^*$ and SkP, since
they are characterized by different values of the
compression modulus in symmetric nuclear matter,
namely $K_{\infty}$=230, 217,
and 202 MeV, respectively. We also investigate the
effect of
different types of pairing forces on the ISGMR in Cd, Sn and Pb
isotopes. The calculated peak energies and the strength
distributions of ISGMR are compared with available experimental
data. We find that SkP fails completely to describe the ISGMR
strength distribution for all isotopes due to its low value
of the
nuclear matter incompressibility, namely $K_{\infty}$=202 MeV.
On the other
hand, the SLy5 parameter set, supplemented by
an appropriate pairing
interaction, gives a reasonable description
of the ISGMR in Cd and Pb isotopes. A better
description of ISGMR in Sn isotopes is achieved by
the SkM$^*$ interaction, that has a somewhat softer
value of the nuclear incompressibility.
\end{abstract}


\maketitle

\section{Introduction}

The compression modulus associated with the
nuclear many-body systems plays an important role
in the description of the structure of finite nuclei,
in the dynamics of heavy-ion collisions,
and in the physics of neutron stars and core-collapse
supernovae \cite{Harakeh,Blai80}. For more than thirty years,
much effort has been done to deduce the value of the
nuclear matter incompressibility both theoretically and experimentally.
The measurements of the compression modes, such as the
isoscalar giant monopole resonance (ISGMR) in finite heavy nuclei,
have been the best tool so far to determine the value of the nuclear matter
incompressibility K$_\infty$. The analysis has mostly, but not
only, been based on the distribution of the ISGMR strength in $^{208}$Pb.
The main results have been reviewed in Ref. \cite{shlomo}.
The extracted values of K$_\infty$ are somewhat model-dependent
but the accepted value from $^{208}$Pb is 240$\pm$20 MeV.
In general, mean field models, either non-relativistic or relativistic,
have been used to extract the value of the nuclear matter incompressibility.
Widely used non-relativistic Skyrme models give a value for the nuclear matter
incompressibility around 230 MeV. It had been claimed that relativistic
mean field models are characterized by larger values of the nuclear matter
incompressibility (around 250 MeV). However, using new fitted Skyrme forces
with different density dependence, the authors of Ref. \cite{colo04} have pointed
out that forces with K$_\infty$ = 250 MeV can also reproduce the ISGMR
experimental data of $^{208}$Pb very well. So, the residual
model dependence in the extracted value of K$_\infty$ is attributed
to the fact that the distribution of ISGMR in
$^{208}$Pb is also sensitive to the density dependence of the
symmetry energy \cite{Vre03,colo04,Pie04}.

Recently, the distribution of ISGMR strength in Cd, Sn and Pb isotopes has
been measured at the Research Center for Nuclear Physics (RCNP) at
Osaka University \cite{Garg11,Li07,Li10,Fuji11}. These data raise
a further question on the nuclear matter incompressibility. Namely,
the Skyrme effective interactions with K$_\infty \sim$ 230 MeV
and the relativistic mean field (RMF) models having K$_\infty \sim$ 270 MeV
can reproduce the experimental ISGMR distribution in
$^{90}$Zr and $^{208}$Pb very well. However, the same models
overestimate the centroid energies of the ISGMR in Sn isotopes \cite{Li07,Li08,Pie07,Nik08}.
This discrepancy shows the fact that the observed ISGMR
in Sn isotopes is softer than those in $^{90}$Zr and $^{208}$Pb,
and this might  be related to our incomplete understanding
of surface, asymmetry and pairing contributions to the
incompressibility of finite nuclei.

Skyrme models have been used in Ref. \cite{Sagawa07} to
investigate the correlation between the asymmetry
contribution to the incompressibility $K_\tau$ and
$K_\infty$, and to validate the extraction of $K_\tau$
from the Sn data. Models having $K_\tau$ and $K_\infty$
compatible with the Sn data, and built within the RMF
framework, are shown to significantly underestimate the
the distribution of strength in $^{208}$Pb \cite{Pie09}.
By calculating the ISGMR in nuclei with large neutron
excess, the authors of Ref. \cite{Cen10}
concluded that
the incompressibility of neutron-rich matter
is still an important open problem.
It has also been pointed out that superfluidity
may have a sizable effect on the incompressibility of
nuclear matter and finite nuclei \cite{Khan09,Khan091}.
This conclusion has been drawn by exploiting
constrained Hartree-Fock
or Hartree-Fock Bogoliubov calculations
\cite{colo04,Bohi79,colo09} in order to determine
the inverse energy-weighted sum rule $m_{-1}$,
and by using this together with the energy-weighted
sum rule $m_{1}$ to define the ISGMR centroid as
$E_{\rm ISGMR}=\sqrt{m_1/m_{-1}}$. It should be
stressed that the effect of pairing on the ISGMR,
within the self-consistent quasi-particle Random
Phase Approximation (QRPA) on top of HFB, has been
highlighted in Ref. \cite{Li08}: in this work it has been
shown that the inclusion of pairing reduces, for some extent,
the discrepancy between the values of $K_\infty$
extracted from $^{208}$Pb and Sn isotopes
data, respectively.

A different type of calculation was performed in
Ref. \cite{Tse09} to describe the ISGMR strength
distribution in Sn isotopes and $^{208}$Pb. The
theoretical models used are the QRPA and the
quasi-particle time blocking approximation (QTBA),
that includes quasi-particle-phonon coupling.
Also in this case, a satisfactory description of
$^{208}$Pb and Sn isotopes at the same time has
not been achieved. This calculation is not
a fully self-consistent one.

In this work, we employ the Skyrme QRPA approach on top of
HF-Bardeen-Cooper-Schrieffer (HF-BCS), to study the
ISGMR strength distribution in Cd, Sn and Pb isotopes.
This method allows systematic and fully self-consistent
calculations for ISGMR. All terms of the interaction, including
the one-body and two-body spin-orbit and Coulomb parts, are included when the
ground-state mean-field and the residual interaction are
evaluated. Three Skyrme parameter sets, SLy5 \cite{Cha98},
SkM$^*$ \cite{Bar82}, and SkP \cite{Dob84} are used in the
calculations. These Skyrme interactions display different values
for the nuclear matter incompressibility.
We also compare the effect of volume, surface, and mixed pairing
interactions on the ISGMR properties in Cd, Sn and Pb isotopes.
This paper is organized as follows. We will briefly report the
main features of our Skyrme HF-BCS plus QRPA model in Sec. II.
The results for ISGMR in Cd, Sn and Pb isotopes are discussed
and compared with available experimental data in Sec. III.
Section IV is devoted to the summary and discusses the
perspectives for future work.

\section{Theoretical Models}

The Skyrme interaction is quite successful in the description
of nuclear properties both of ground states and excited states.
As mentioned in the previous Section, we will use the QRPA approach
on top of HF-BCS for our theoretical investigation.

Firstly we solve the Skyrme HF-BCS equation for the ground state
in coordinate space. The radial mesh on which the equations are
solved extends up to 18 fm,
and the mesh size is 0.1 fm.
For all the nuclei under study,
this radial mesh is large enough so that the results are stable.
The pairing correlations are generated by a density-dependent zero-range force,
\begin{equation}
V_{pair}(\mathbf{r}_1,\mathbf{r}_2)=V_0\left[1-\eta\left(\frac{\rho(\mathbf{r})}{\rho_0}\right)\right]\delta(\mathbf{r}_1-\mathbf{r}_2),
\label{eq:pair}
\end{equation}
where $\rho(\mathbf{r})$ is the particle density in coordinate space
and $\rho_0$=0.16 fm$^{-3}$ is the nuclear saturation density.
The value of $\eta$ is taken as 0, 0.5, or
1 for the volume, mixed, or surface pairing interactions,
respectively. The pairing window, that is, the states
that are taken into account for the solution of the BCS
equations, includes five unoccupied orbitals above the last occupied
level in the HF approximation. In this space, the pairing strength $V_0$
is fixed by fitting the pairing gap extracted from experimental
data of odd-even mass difference by using the five point formula.
The values of $V_0$ that we have obtained are displayed
in Table I. They reproduce well the gap associated with
one typical nucleus for each isotope chain; in particular,
we have chosen to reproduce
the empirical pairing gaps of
$^{112}$Cd ($\Delta_n$=1.334 MeV), $^{120}$Sn
($\Delta_n$=1.321 MeV), and $^{204}$Pb ($\Delta_n$=0.841 MeV).
For Sn and Pb isotopes, there is only neutron pairing
because these nuclei have closed proton shells associated with
Z = 50 and 82,  respectively.
For Cd isotopes, the proton pairing
also exists in principle because the proton number 48 is not
a magic number.
However, we have found numerically
that the effect of proton pairing in Cd isotopes on
the ISGMR strength distribution
is very small (of the order of tenths of keV).
So in our calculation we use the
filling approximation for the proton 1$g_{9/2}$ state in Cd isotopes.
The pairing strengths $V_0$
obtained by fitting the empirical gaps are
different
for each Skyrme parameter set and for each pairing model.

\begin{table*}
\caption{The pairing strength $V_0$ for various types of pairing
interactions defined in Eq. \eqref{eq:pair}, in units of MeV
fm$^{3}$. For details see the text.}
\begin{ruledtabular}
\begin{tabular}{ccccc}

           &          &   volume  &  surface  &   mixed    \\
           \hline
           &  SLy5    & 261       &  738      &   388      \\
$^{112}$Cd &  SkM$^*$ & 230       &  675      &   342      \\
           &  SkP     & 215       &  692      &   328      \\
\hline
           &  SLy5    & 218       &  645      &   325      \\
$^{120}$Sn &  SkM$^*$ & 255       &  725      &   381      \\
           &  SkP     & 213       &  688      &   328      \\
\hline
           &  SLy5    & 265       &  875      &   409      \\
$^{204}$Pb &  SkM$^*$ & 255       &  863      &   392      \\
           &  SkP     & 211       &  771      &   335      \\
\end{tabular}
\end{ruledtabular}
\end{table*}

We now provide few details about the QRPA calculations.
The single-particle continuum is discretized by setting
the nuclei in a spherical box of radius equal to 18 fm.
For every value of the quantum numbers ($l$, $j$) associated
with the single-particle states,
we include in the QRPA model space the unoccupied
states up to
the maximum number of nodes given by
$n_{max}$ = $n_{\rm last}$+12, where $n_{\rm last}$ is the number
of nodes of the last occupied state with a given ($l$, $j$).
The convergence of the calculated results is checked by
looking at the results for the energy
and the strength of the ISGMR.
The QRPA matrix equation
having good
angular momentum and parity J$^\pi$ is given by
\bea\label{RPA}
\left( \begin{array}{cc}
  A & B \\
B^* & A^* \end{array}  \right)
\left( \begin{array}{c}
X^n \\
Y^n  \end{array} \right) =\hbar\omega_n
\left( \begin{array}{cc}
1 & 0\\
0 & -1 \end{array}  \right)
\left( \begin{array}{c}
X^n \\
Y^n  \end{array} \right), \eea
where $\hbar \omega_n$ is the energy of the $n$-th
QRPA state and  X$^n$, Y$^n$ are the corresponding
forward and backward amplitudes, respectively. The explicit forms of the matrices A and B are given elsewhere
\cite{Rowe70, Ring80, Sev02, colo11}.
The p-h matrix elements are derived from the Skyrme
energy density functional including all the terms such as the two-body
spin-orbit and two-body Coulomb interactions. We
should mention that several previous works
devoted to the study of the ISGMR strength in Sn
isotopes
are not fully
self-consistent \cite{Sagawa07,Li08,Tse09} since the
spin-orbit interaction is not taken into account
in the residual interaction.
In Ref. \cite{Sil06}, the authors have discussed the self-consistency
violation in HF-RPA calculations for nuclear giant
resonances: they have shown, for example,
that the two-body spin-orbit interaction gives a slight repulsive
contribution to the ISGMR strength in light nuclei, whereas, for
medium and heavy nuclei, it produces an attractive effect on the
ISGMR strength, so that the centroid energies are pushed
downward by about 0.6 MeV.

After solving the QRPA equations, various moments of the
strength distributions can be obtained by means of the equation
\begin{equation}
m_k=\int E^kS(E)dE,
\end{equation}
where $S(E)=\sum_n|\langle 0|\hat{F}|n\rangle|^2\delta(E-E_n)$
is the strength function associated with the monopole operator
\begin{equation}
\hat{F}=\sum_i r^2_i,
\end{equation}
The constrained energy E$_{con}$, the centroid energy E$_{cen}$, and
the scaling energy E$_{s}$ of the resonance are then defined as
\begin{equation}
E_{con}=\sqrt{\frac{m_1}{m_{-1}}},~~~E_{cen}
=\frac{m_1}{m_0},~~~E_{s}=\sqrt{\frac{m_3}{m_{1}}},
\end{equation}
respectively.

\section{ISGMR Results and Discussion}

In order to check how much the pairing correlations affect
the ISGMR strength distributions,
we show in Fig. 1 the ISGMR strength distributions in
$^{110}$Cd and $^{120}$Sn calculated by
using different pairing models. The effective force
SkM* is adopted in
the particle-hole channel. For the pairing channel, we take the
surface, mixed and volume pairing interactions and compare
the results with those obtained within RPA (no pairing) and
the filling approximation. We find that the surface pairing and
the filling approximation give almost identical results. On the
other hand, the volume and the mixed pairing predict slightly lower
peak energies although the difference is rather small.
The ISGMR centroid energy in $^{110}$Cd ($^{120}$Sn) are 16.38 (15.80),
16.33 (15.86), 16.10 (15.69), 16.18 (15.76) MeV, when
calculated by employing either the filling approximation,
or the surface, or the volume, or the mixed pairing interactions,
respectively. The maximum difference due to different
pairing models is 280 (110) keV in
$^{110}$Cd ($^{120}$Sn).

\begin{figure*}[hbt]
\includegraphics[width=0.47\textwidth]{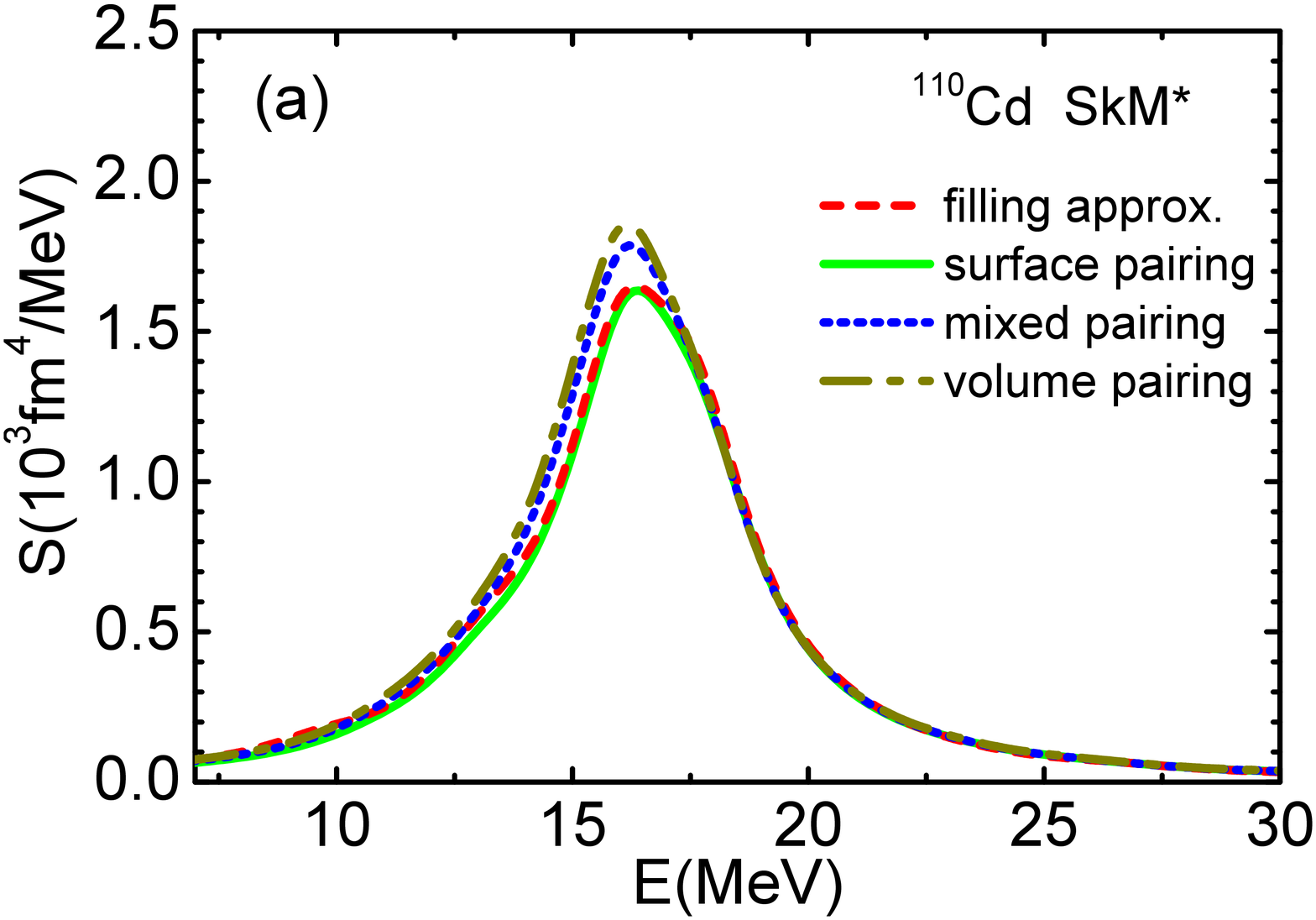}
\includegraphics[width=0.47\textwidth]{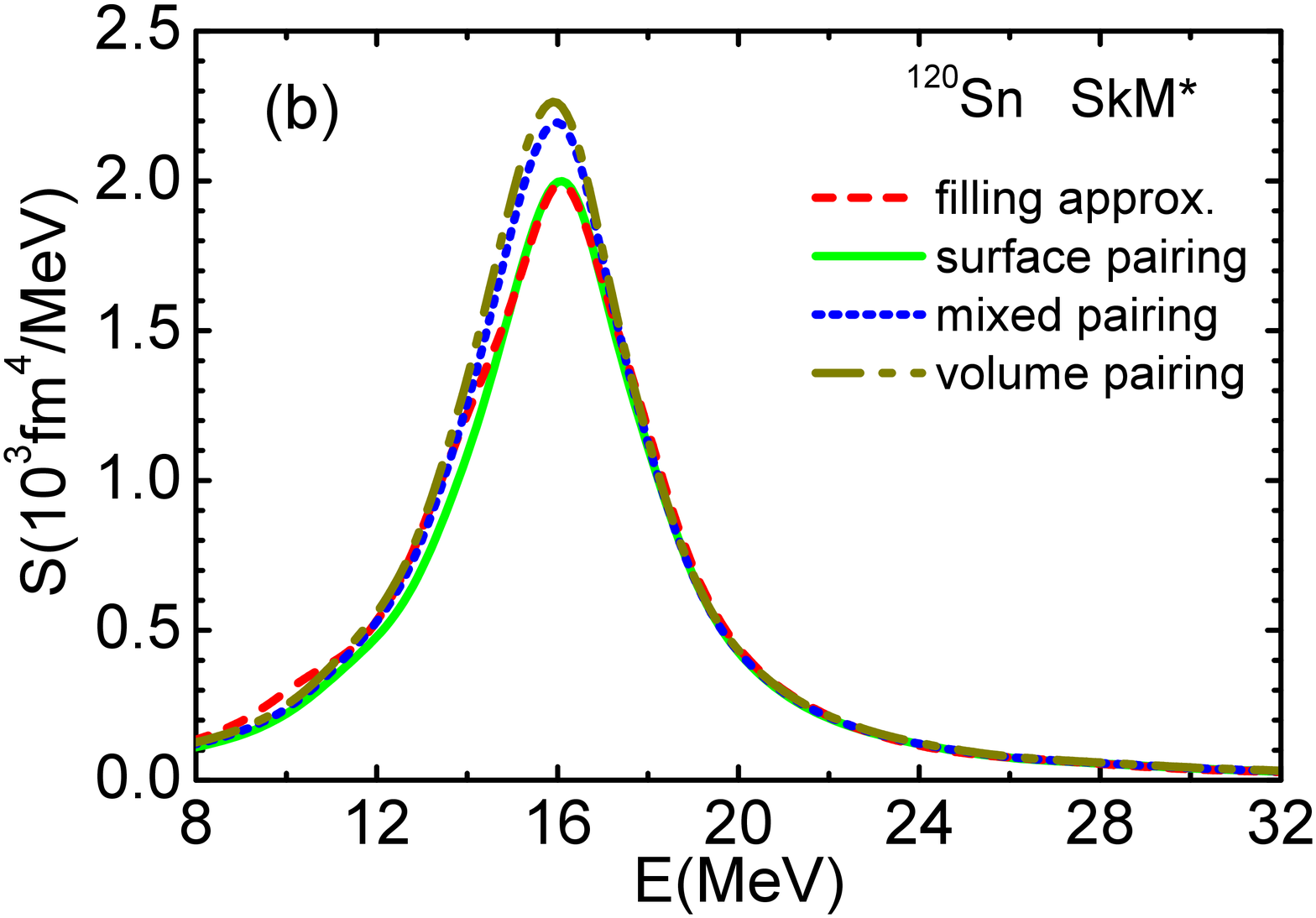}
\vglue -1.3cm
\caption{The ISGMR strength distribution in (a) $^{110}$Cd
and (b) $^{120}$Sn,
calculated by using either the filling approximation,
or the volume, or the surface, or the mixed pairing forces.
The SkM* force is adopted in the p-h channel.
} \label{Fig.1}
\end{figure*}

\subsection{Cd Isotopes}

Very recently, measurements of the ISGMR strength distributions
in Cd isotopes were performed at RCNP, Osaka University \cite{Garg11}.
In keeping with the fact that it is difficult to reproduce
equally well Pb and Sn isotopes with a unique Skyrme force, we
would like to see what are the results that these forces provide
for the strength distributions in
Cd isotopes.
To address this question, we have performed the calculations
for the ISGMR strength in the Cd isotopes with the three
different aforementioned Skyrme interactions together with
various pairing models. Figure
2 displays the QRPA results for the ISGMR strength distribution
in $^{106-116}$Cd,
calculated by using the
SkP (dotted line), SkM* (dashed line), and SLy5 (solid line)
interactions, respectively.
The pairing force adopted in Fig. 2 is the mixed pairing
interaction. For all nuclei,
the strength distributions are concentrated in a single peak
around 16 MeV, which
exhausts almost all the energy weighted sum rule. However,
the location of the peak found with each Skyrme interaction
is slightly different.
The SkP interaction predicts lowest peaks
while the SLy5 interaction gives peaks at the highest energies.
The peaks obtained by using the SkM* interaction stays in
the middle between the other two cases.
As it is known from previous studies, the relative position
of the peaks
is governed by the nuclear matter incompressibility
associated with each effective interaction.

\begin{figure*}[hbt]
\includegraphics[width=0.47\textwidth]{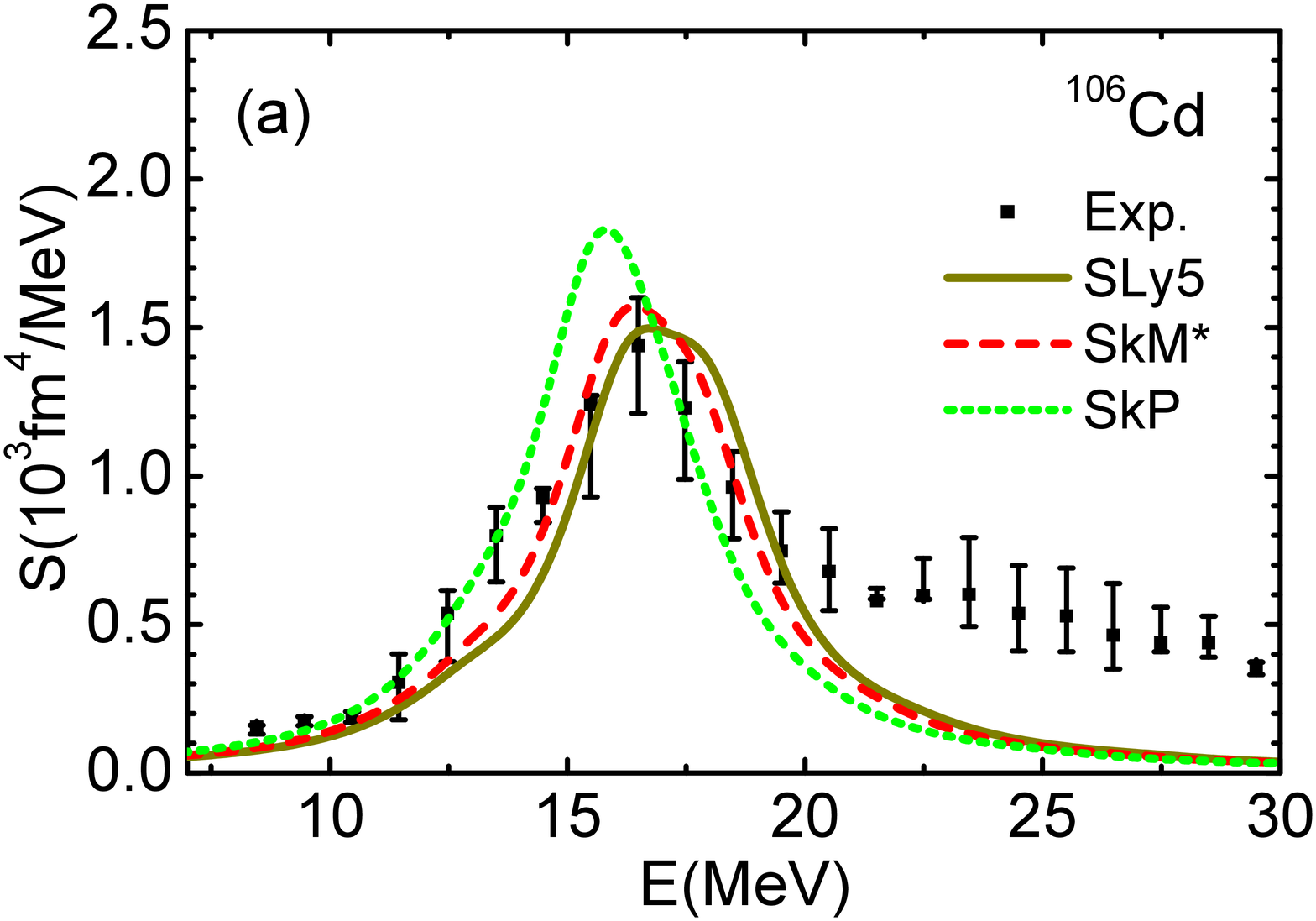}
\includegraphics[width=0.47\textwidth]{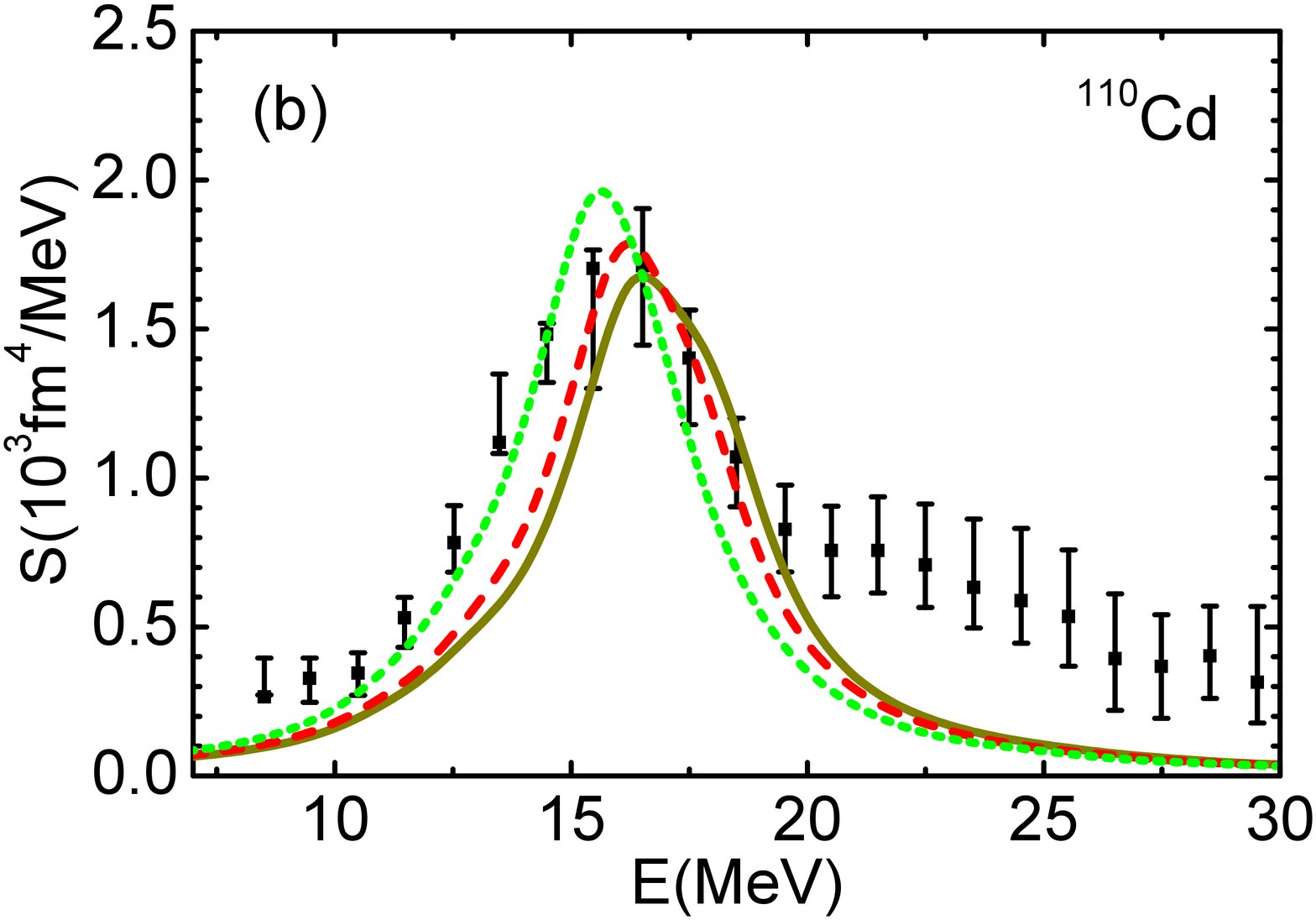}
\vglue -2cm
\includegraphics[width=0.47\textwidth]{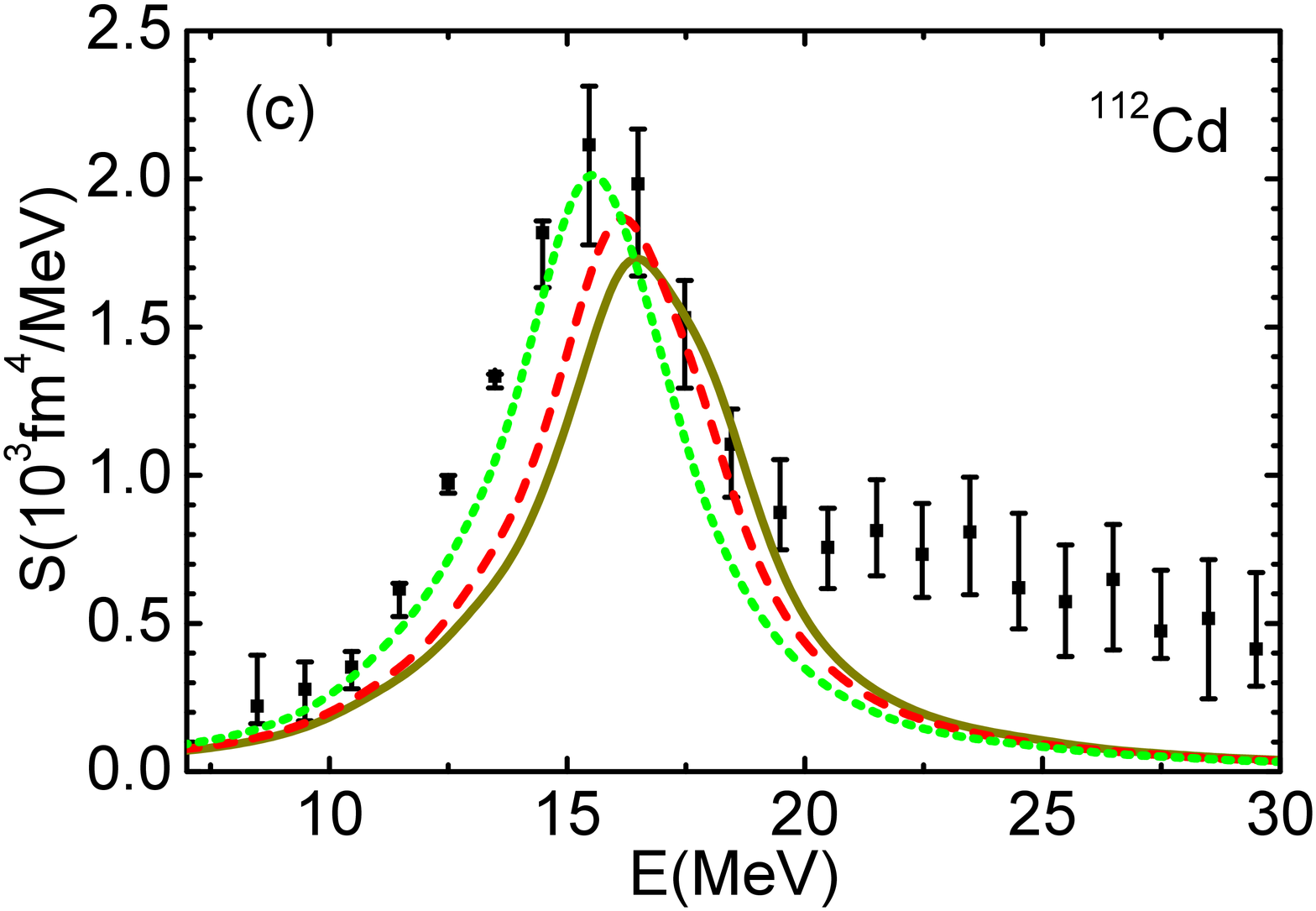}
\includegraphics[width=0.47\textwidth]{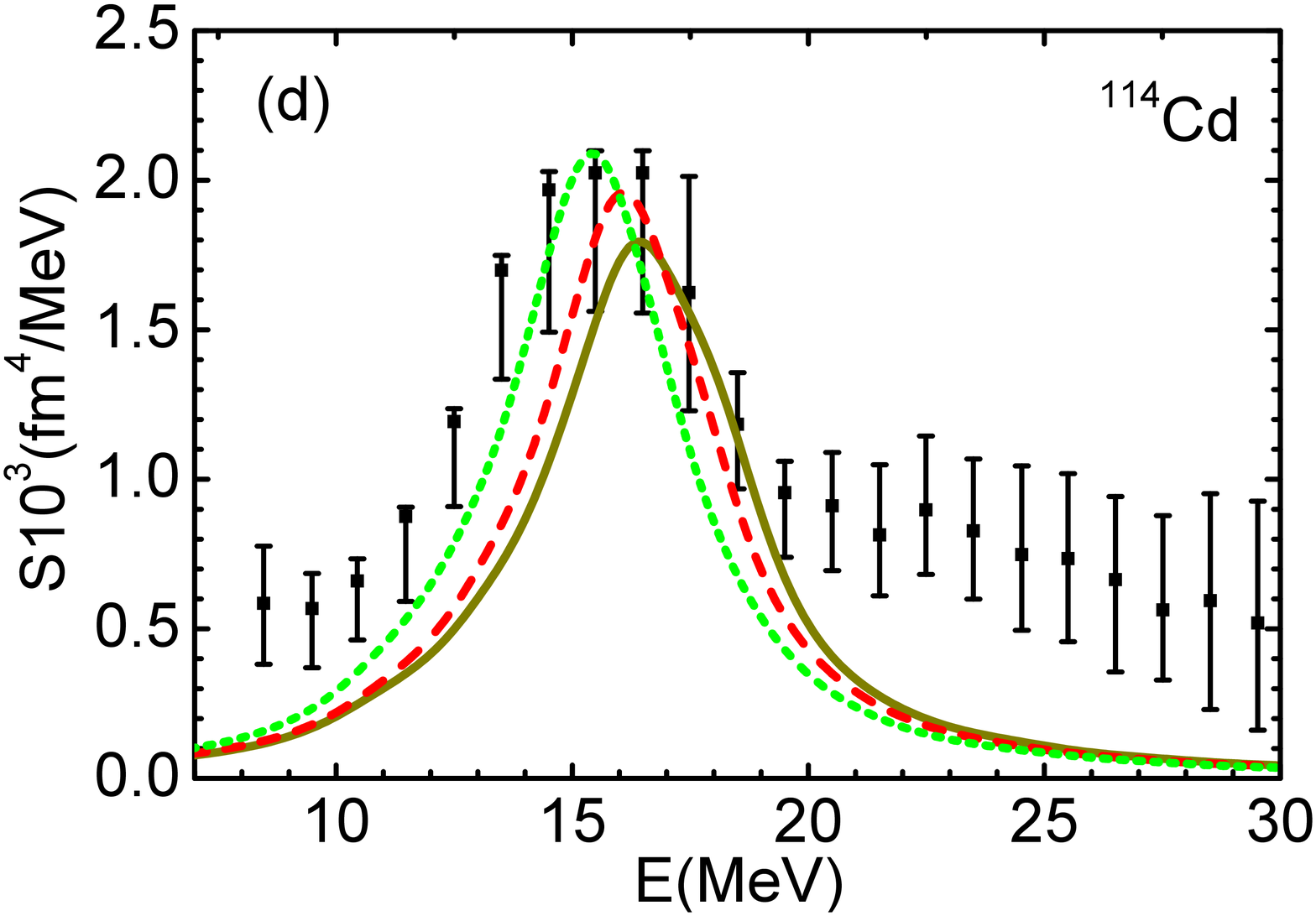}
\vglue -2cm
\includegraphics[width=0.47\textwidth]{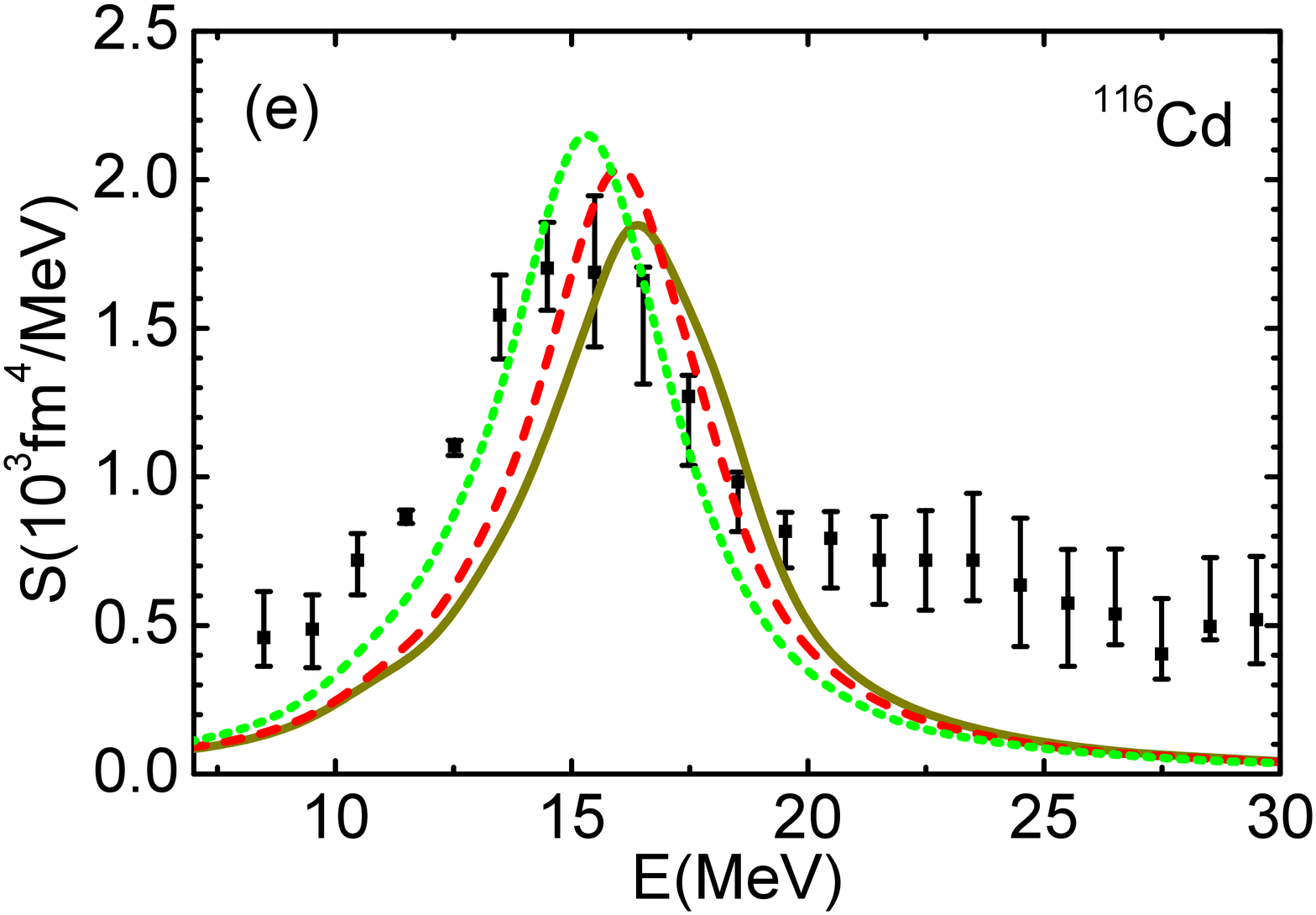}
\vglue -2cm
\caption{The calculated ISGMR strength distribution in
$^{106-116}$Cd are compared with the experimental
data obtained at RCNP \cite{Garg11}.
The SLy5 (solid line), SkM* (dashed line) and SkP (dotted line)
forces are adopted in the calculations whose results are
shown here, together with the mixed pairing interaction.
} \label{Fig.2}
\end{figure*}

\begin{figure*}[hbt]
\includegraphics[width=0.47\textwidth]{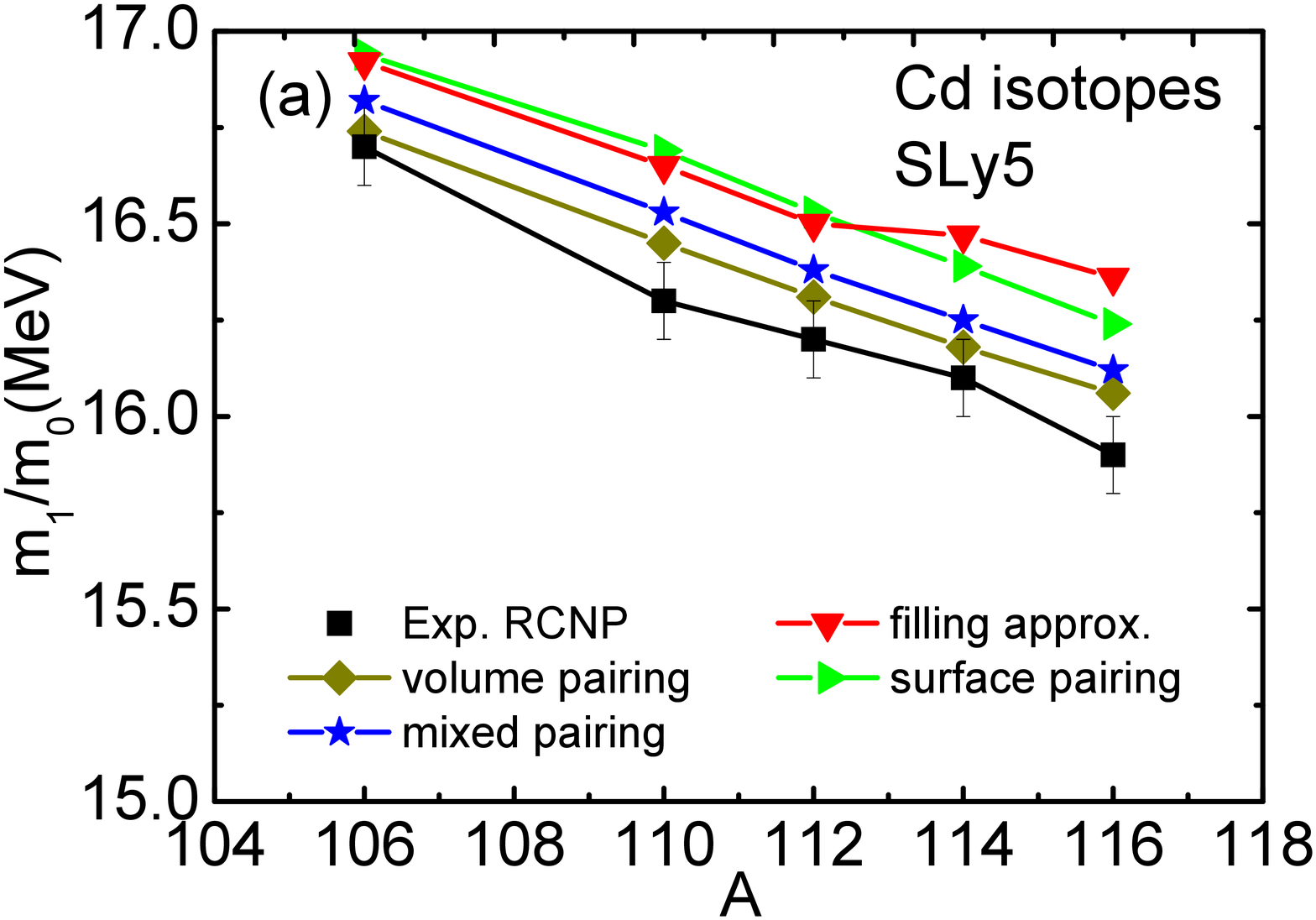}
\includegraphics[width=0.47\textwidth]{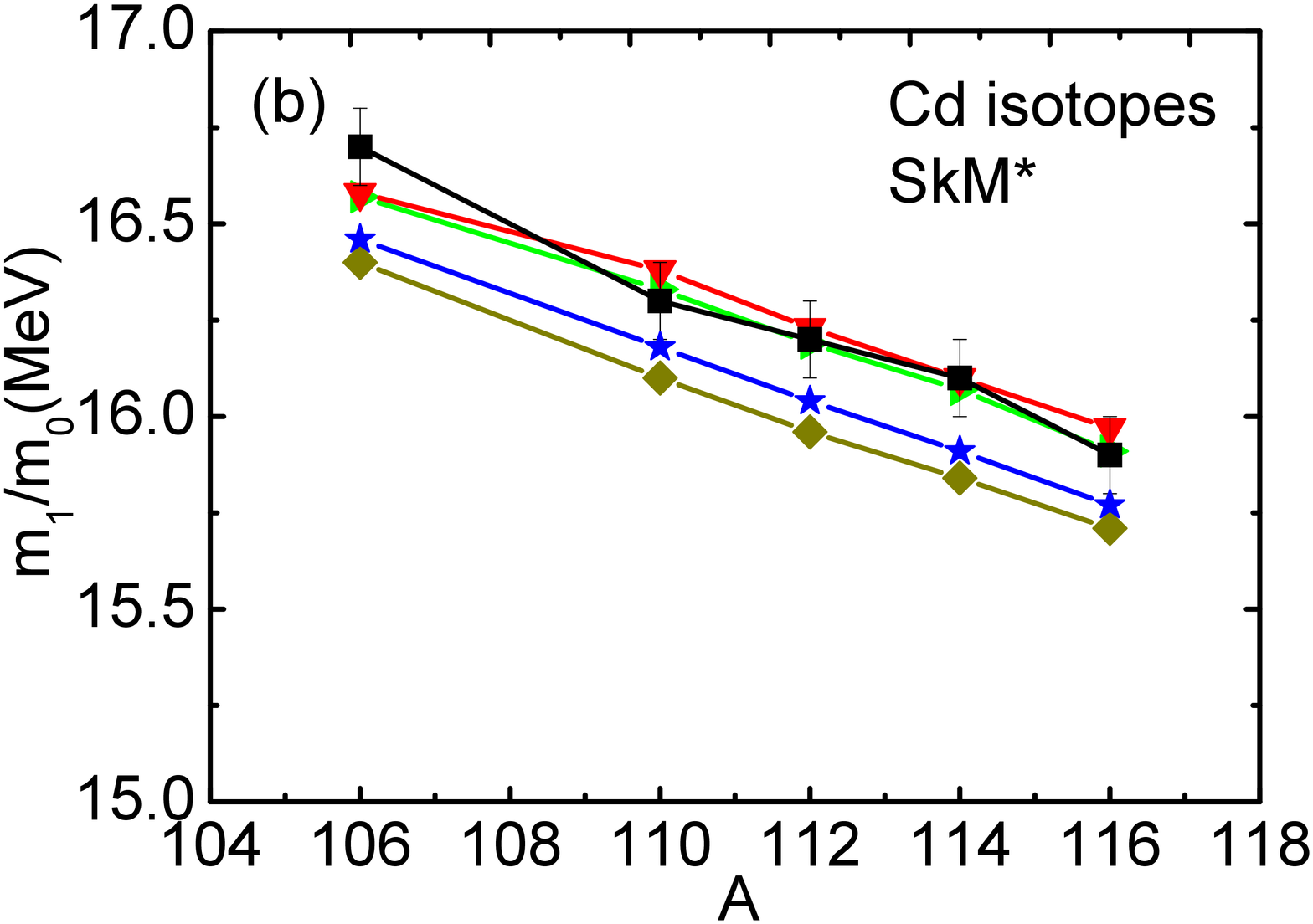}
\vglue -2cm
\includegraphics[width=0.47\textwidth]{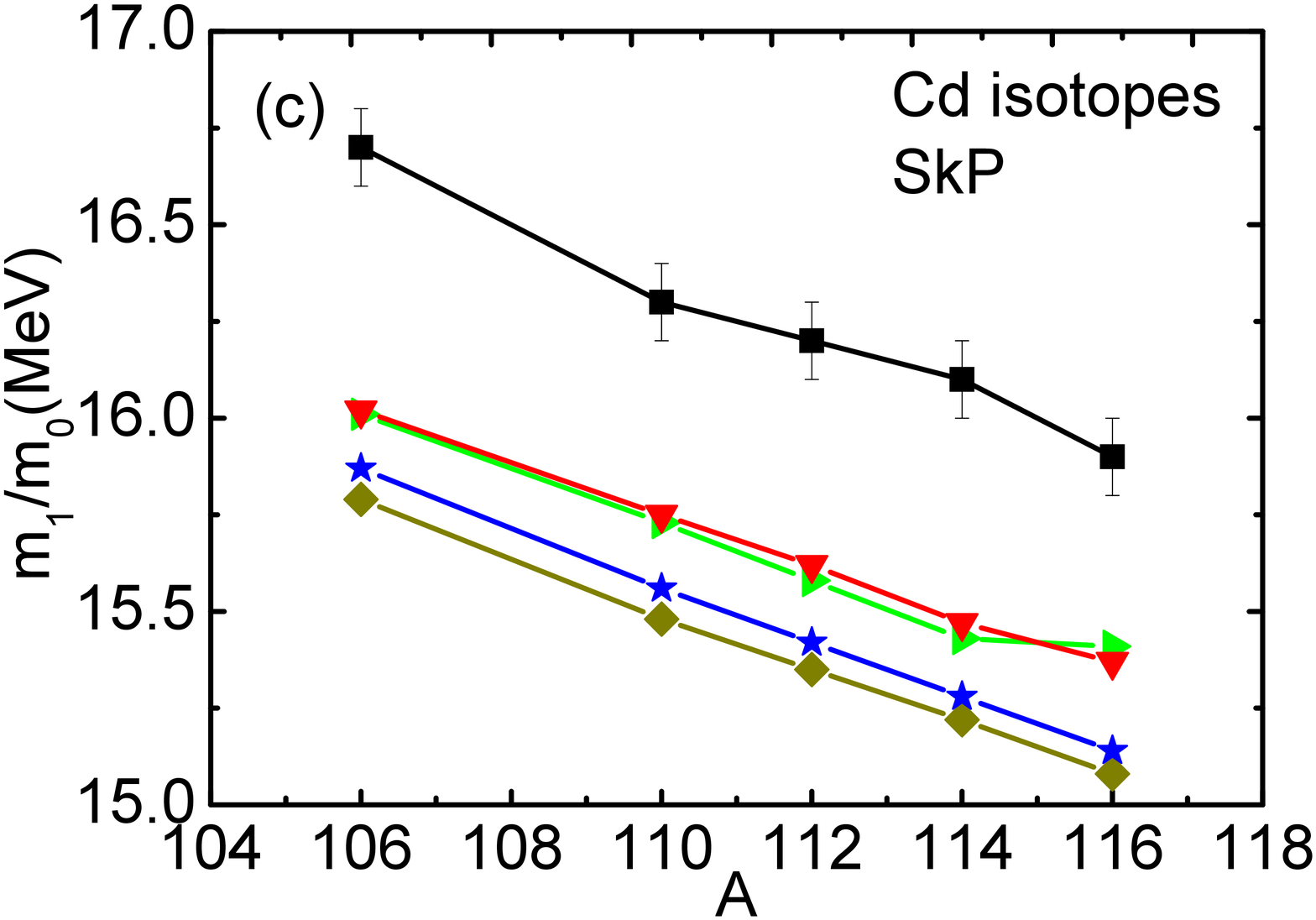}
\vglue -2cm
\caption{The calculated ISGMR centroid energies in the
even-even $^{106-116}$Cd isotopes are compared with
the experimental data obtained from Ref. \cite{Garg11}.
The forces SLy5 (a), SkM* (b) and SkP (c) are adopted
in the present calculations together with either the filling
approximation, or the volume, or the surface, or the
mixed pairing interactions, respectively.} \label{Fig.3}
\end{figure*}

\begin{table*}
\caption{The calculated ISGMR constrained energies ($E_{con}$),
centroid energies ($E_{cen}$), and scaling energies ($E_{s}$)
in even-even $^{106-116}$Cd isotopes are compared with
the experimental data. The theoretical results are obtained
in the interval between 10.5 and 20.5 MeV, by using the
SKP, SKM*, and SLy5 parameter sets together with the mixed
pairing interaction. The experimental data are from
Ref. \cite{Garg11}. The values in parenthesis are the
differences between the
theoretical values and the experimental data. Units are MeV.}
\begin{ruledtabular}
\begin{tabular}{ccccccccccccc}

            &     &  Exp.    &   SkP  &  SkM*  &   SLy5      \\
\hline
      &$^{106}$Cd & 16.5$\pm$0.2 & 15.81(-0.69) & 16.40(-0.1) & 16.75(0.25) \\
      &$^{110}$Cd & 16.1$\pm$0.2 & 15.50(-0.60) & 16.11(0.01) & 16.46(0.36)  \\
$E_{con}=\sqrt{m_1/m_{-1}}$
      &$^{112}$Cd & 16.0$\pm$0.2 & 15.35(-0.65) & 15.97(-0.03) & 16.29(0.29)  \\
      &$^{114}$Cd & 15.8$\pm$0.2 & 15.21(-0.59) & 15.84(0.04) & 16.16(0.36) \\
      &$^{116}$Cd & 15.7$\pm$0.2 & 15.06(-0.64)  & 15.70(0.0) & 16.02(0.32)  \\
\hline
      &$^{106}$Cd & 16.7$\pm$0.1 & 15.87(-0.83) & 16.46(-0.24) & 16.82(0.12) \\
      &$^{110}$Cd & 16.3$\pm$0.1 & 15.56(-0.74) & 16.18(-0.12) & 16.54(0.24) \\
$E_{cen}=m_1/m_0$
      &$^{112}$Cd & 16.2$\pm$0.1 & 15.42(-0.78) & 16.05(-0.15) & 16.39(0.19) \\
      &$^{114}$Cd & 16.1$\pm$0.1 & 15.28(-0.82) & 15.91(-0.19) & 16.25(0.15) \\
      &$^{116}$Cd & 15.9$\pm$0.1 & 15.14(-0.76) & 15.78(-0.12) & 16.12(0.22) \\
\hline
      &$^{106}$Cd & 17.2$\pm$0.3 & 16.02(-1.18) & 16.63(-0.57) & 16.99(-0.21) \\
      &$^{110}$Cd & 16.9$\pm$0.3 & 15.74(-1.16) & 16.37(-0.53) & 16.74(-0.16) \\
$E_{s}=\sqrt{m_3/m_1}$
      &$^{112}$Cd & 16.8$\pm$0.2 & 15.62(-1.18) & 16.24(-0.56) & 16.61(-0.19) \\
      &$^{114}$Cd & 16.7$\pm$0.4 & 15.49(-1.21) & 16.11(-0.59) & 16.49(-0.21) \\
      &$^{116}$Cd & 16.6$\pm$0.3 & 15.35(-1.25) & 15.97(-0.63) & 16.37(-0.23) \\
\end{tabular}
\end{ruledtabular}
\end{table*}

In Fig. 3 we show the theoretical ISGMR centroid energies
in $^{106-116}$Cd obtained by the QRPA calculations.
The results in Figs. 3(a), 3(b) and 3(c)
are obtained by using the SLy5, SkM*, and SkP Skyrme interactions
together with various pairing interactions. Figure 3(a) shows
the ISGMR centroid energies obtained by the SLy5 interaction. With
either the filling approximation or the surface pairing interaction,
the calculated results are about 300 keV systematically larger than
the experimental data.
On the other hand, the calculations with volume and mixed pairing
interaction reproduce
the experimental centroid energies within
150 keV. For the case of the SkM* interaction whose results are
shown in Fig. 3(b), the conclusion is different
from the one obtained by  the SLy5 interaction,
due to the lower value
of the nuclear matter incompressibility (we remind that
K$_{\infty}$ = 217 MeV for SKM*, and that K$_{\infty}$ = 230 MeV
for SLy5). It seems that the filling approximation and the
surface pairing interaction provide a good reproduction of
the experimental centroid energies, whereas the volume and
mixed pairing interactions somewhat underestimate the
experimental data. Finally, it should be expected that the results
obtained with the SkP parameter set underestimate the experimental
ISGMR centroid energies in $^{106-116}$Cd [cf. Fig. 3(c)],
due to the lower value
of the nuclear matter incompressibility associated with this
parameter set, namely K$_{\infty}$ = 201 MeV.
Also in this case the results depend on the choice of the
pairing force: when volume and mixed pairing interactions
are adopted, the results tend to worsen.

Whereas we can clearly confirm from the present results
that the value of the nuclear matter incompressibility
does play a key role in dictating the location of the ISGMR
centroid energy, it is also true that the pairing interaction
lowers the energy of the ISGMR to some extent, typically few
hundreds keV. This qualitative conclusion is the same
that was first found in Ref. \cite{Li08}.
Thus, the pairing interaction cannot be neglected
if one aims to reproduce not only the ISGMR centroid energies in
Cd isotopes, but also,  more generally, in other open-shell
nuclei. It should also be noticed that the slope of
the isotope dependence of the ISGMR
centroid energy is rather well reproduced by all the three
interactions, while the absolute values
are much more sensitive to the choice of the Skyrme parameter
set and of the pairing force.

The various kinds of centroid energies are shown in Table II.
The results that we report here, are obtained with the
mixed pairing interaction for each Skyrme parameter set.
Again, the SkP interaction underestimates all the energies
in Cd isotopes. With the mixed pairing force, SkM* can give
an excellent description of the constrained energies
$E_{con}=\sqrt{m_1/m_{-1}}$ for all nuclei, while a slight
underestimation of the centroid energies $E_{cen}=m_1/m_0$
is produced.
Compared to the other two interactions, in general, SLy5
can provide better results for the energies that are
displayed in Table II.

It is not completely clear, from the experimental point
of view, whether the constrained, centroid or scaling
energies are more suitable to be compared with the
experimental data. However, from what we have just
concluded, it can be stated that the reasonable values
of nuclear incompressibility that can be extracted
from the present Cd data are either the one of
SkM* or SLy5, namely 217 or 230 MeV. This number is
 consistent with the one extracted from
$^{208}$Pb, although on the lower side.

\subsection{Sn Isotopes}

Fig. 4 shows the calculated ISGMR strength distributions
in the even-even $^{112-124}$Sn together with the experimental
data taken from Refs. \cite{Li07,Li10}. The results obtained
with the Skyrme sets SLy5 (solid line), SkM* (dashed line) and
SkP (dotted line), and with the mixed pairing interaction,
are displayed. All three Skyrme interactions give a single
peak around 16 MeV. The interaction dependence of the
peak energy is qualitatively the same as in the case of
Cd isotopes, i.e., the SkP result is the lowest, the SkM*
one lies in the middle and the SLy5 result is found at the
highest energy - in agreement with the associated values
of the incompressibility.

\begin{figure*}[hbt]
\includegraphics[width=0.47\textwidth]{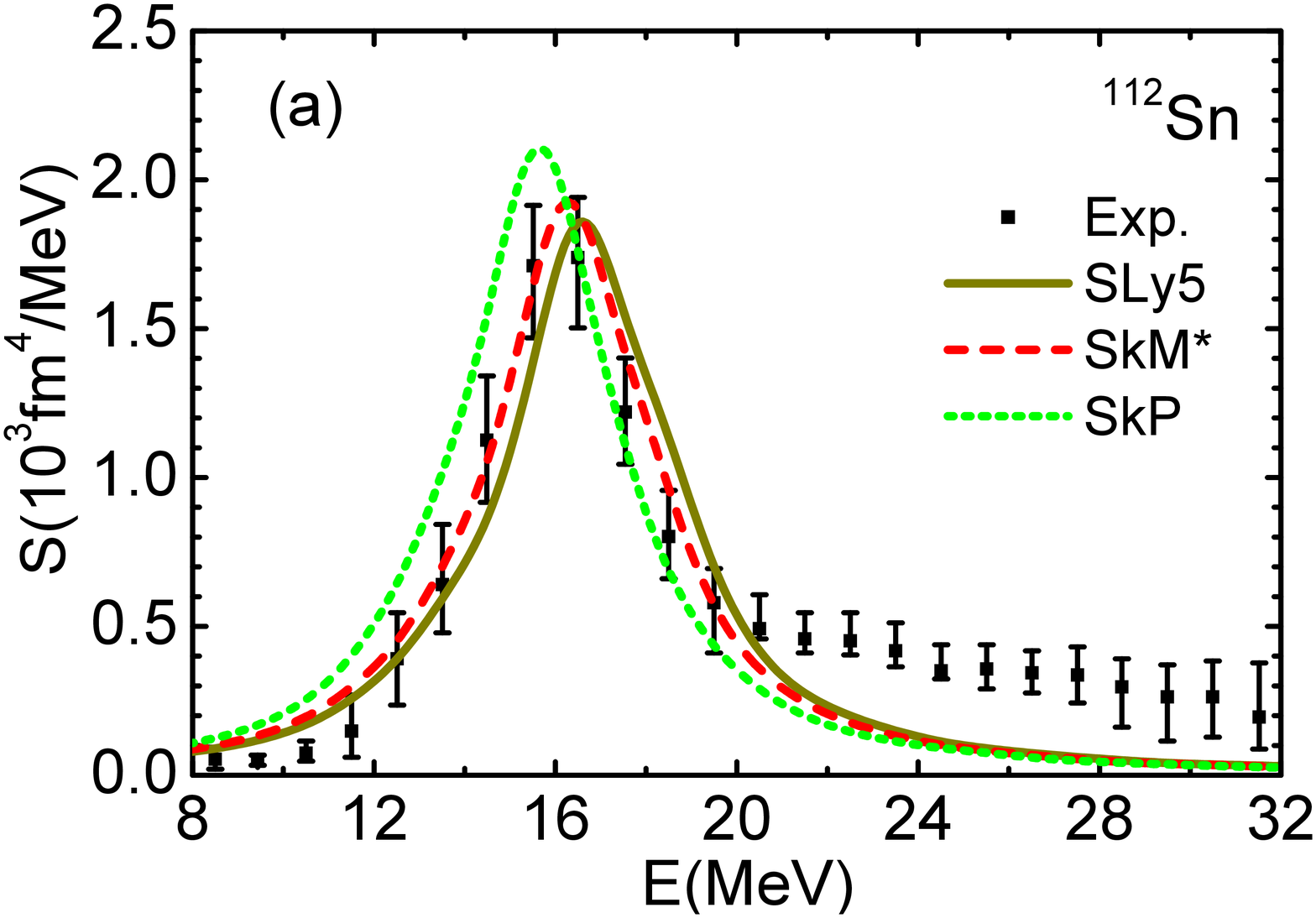}
\includegraphics[width=0.47\textwidth]{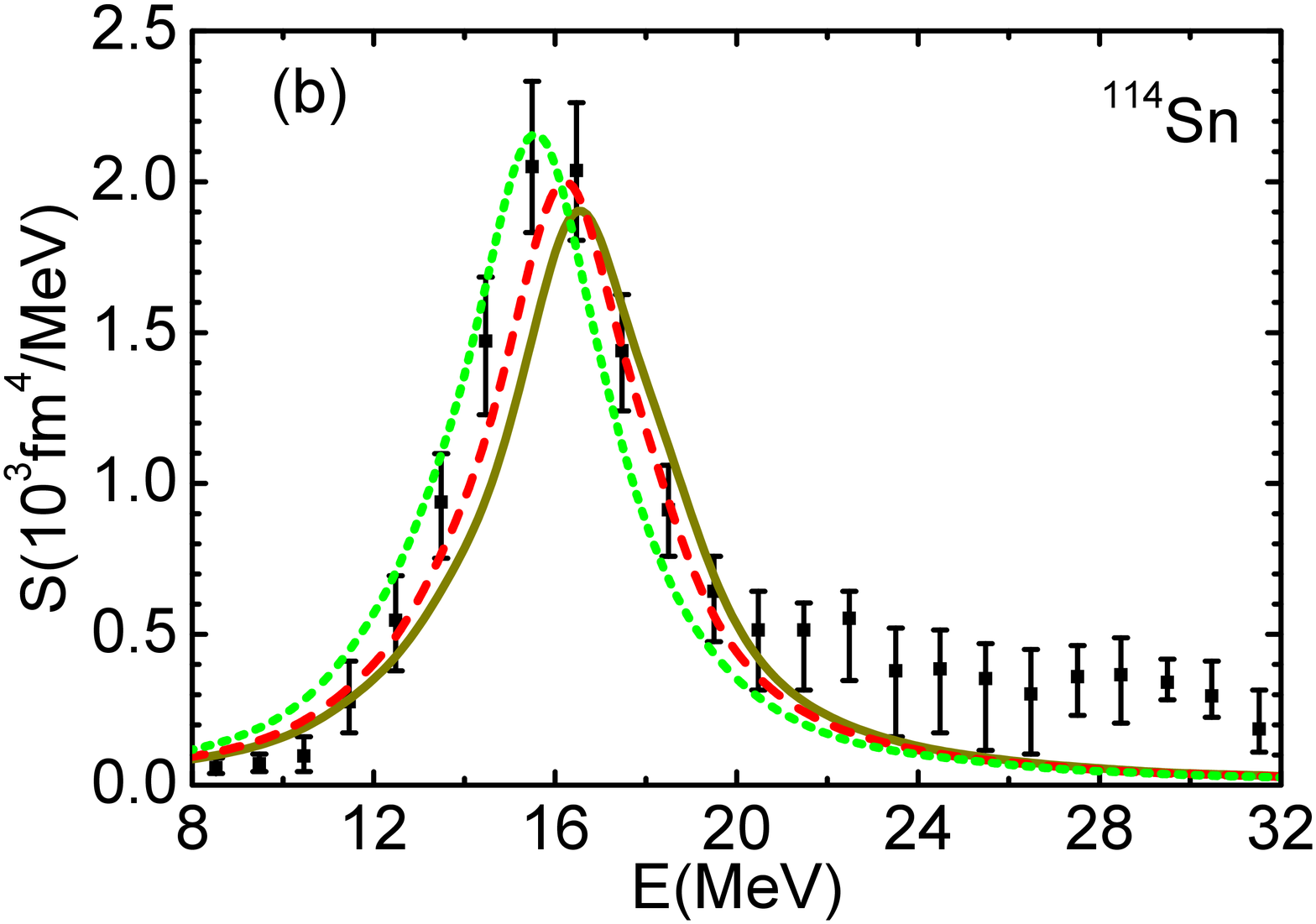}
\vglue -2cm
\includegraphics[width=0.47\textwidth]{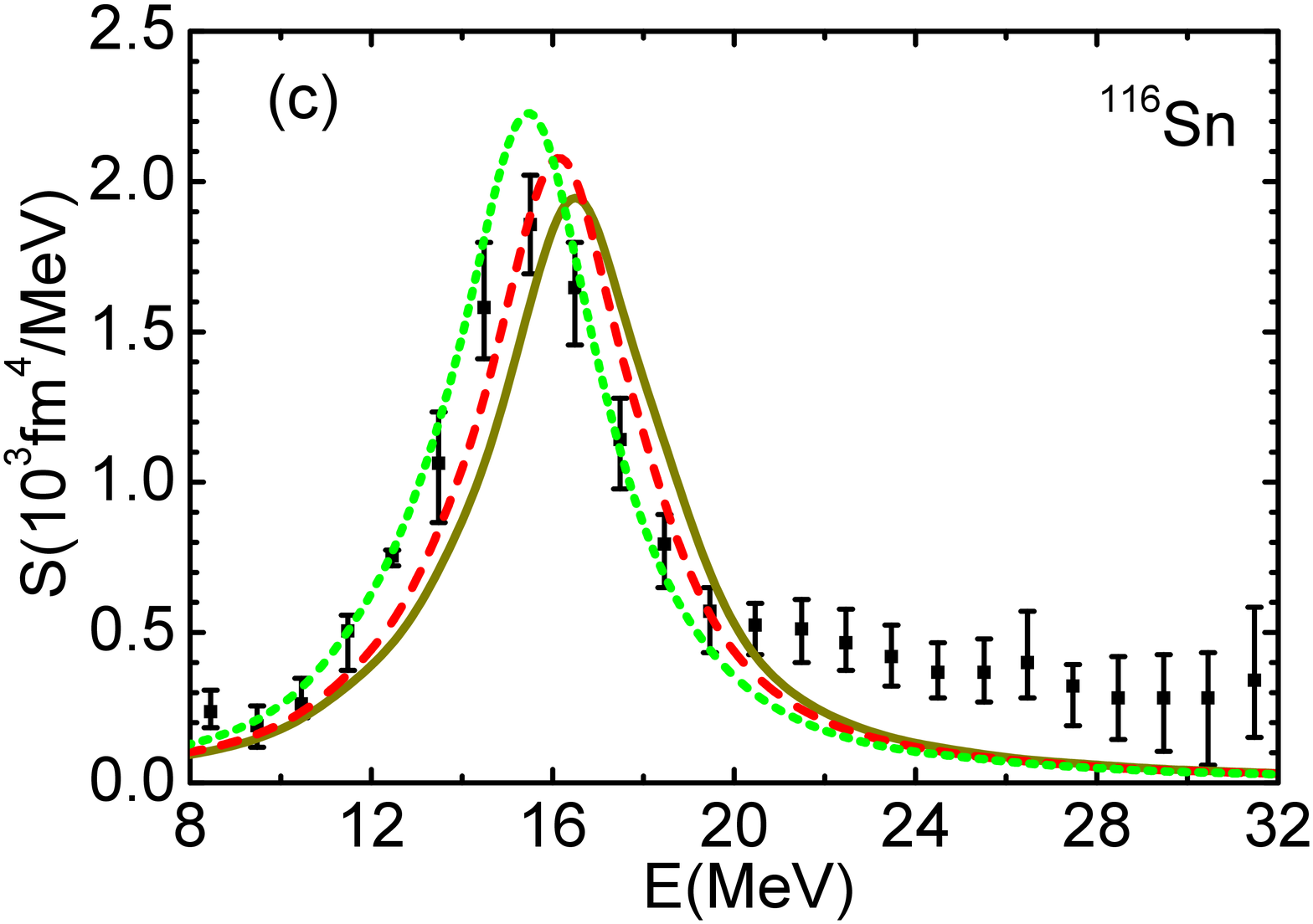}
\includegraphics[width=0.47\textwidth]{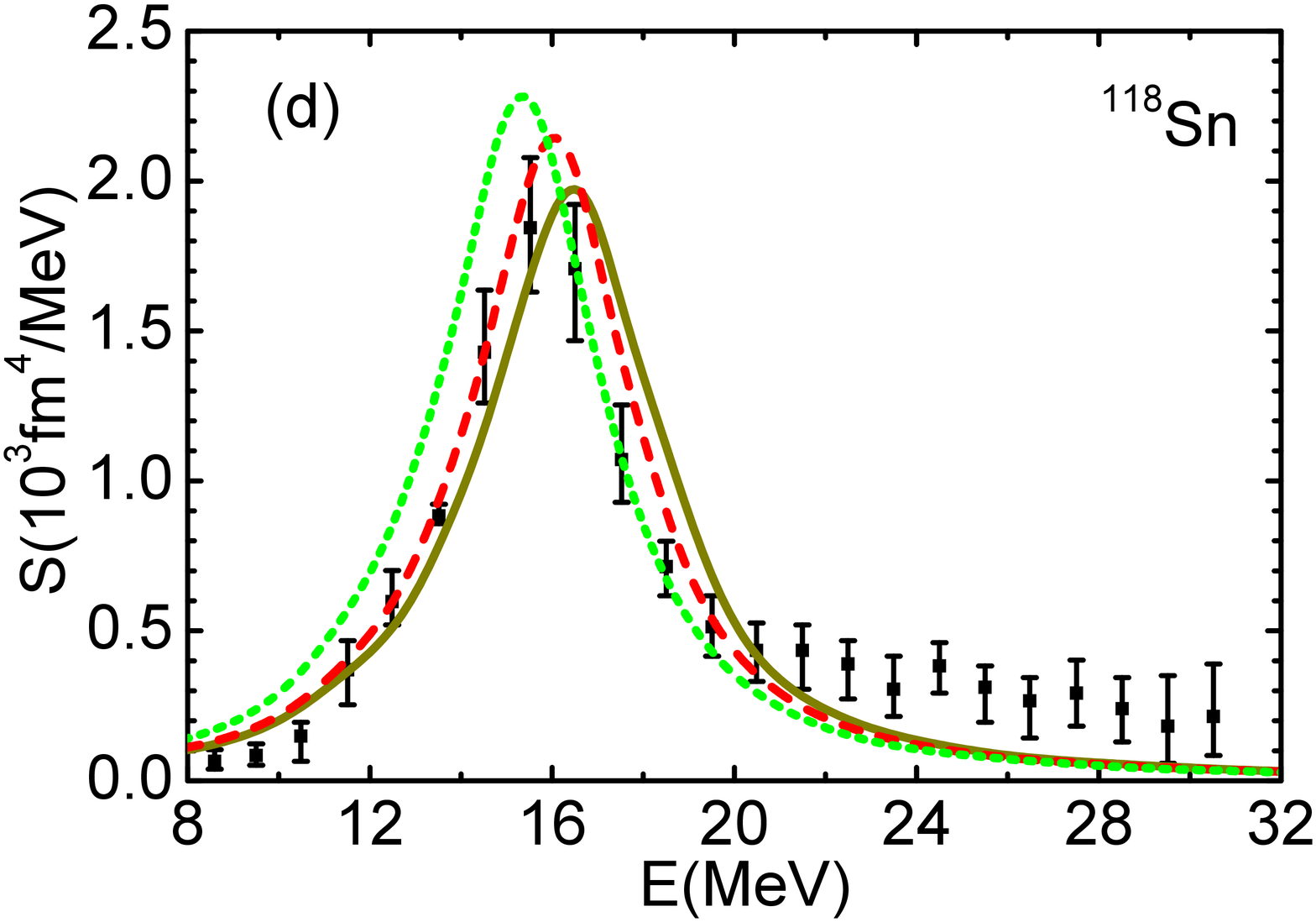}
\vglue -2cm
\includegraphics[width=0.47\textwidth]{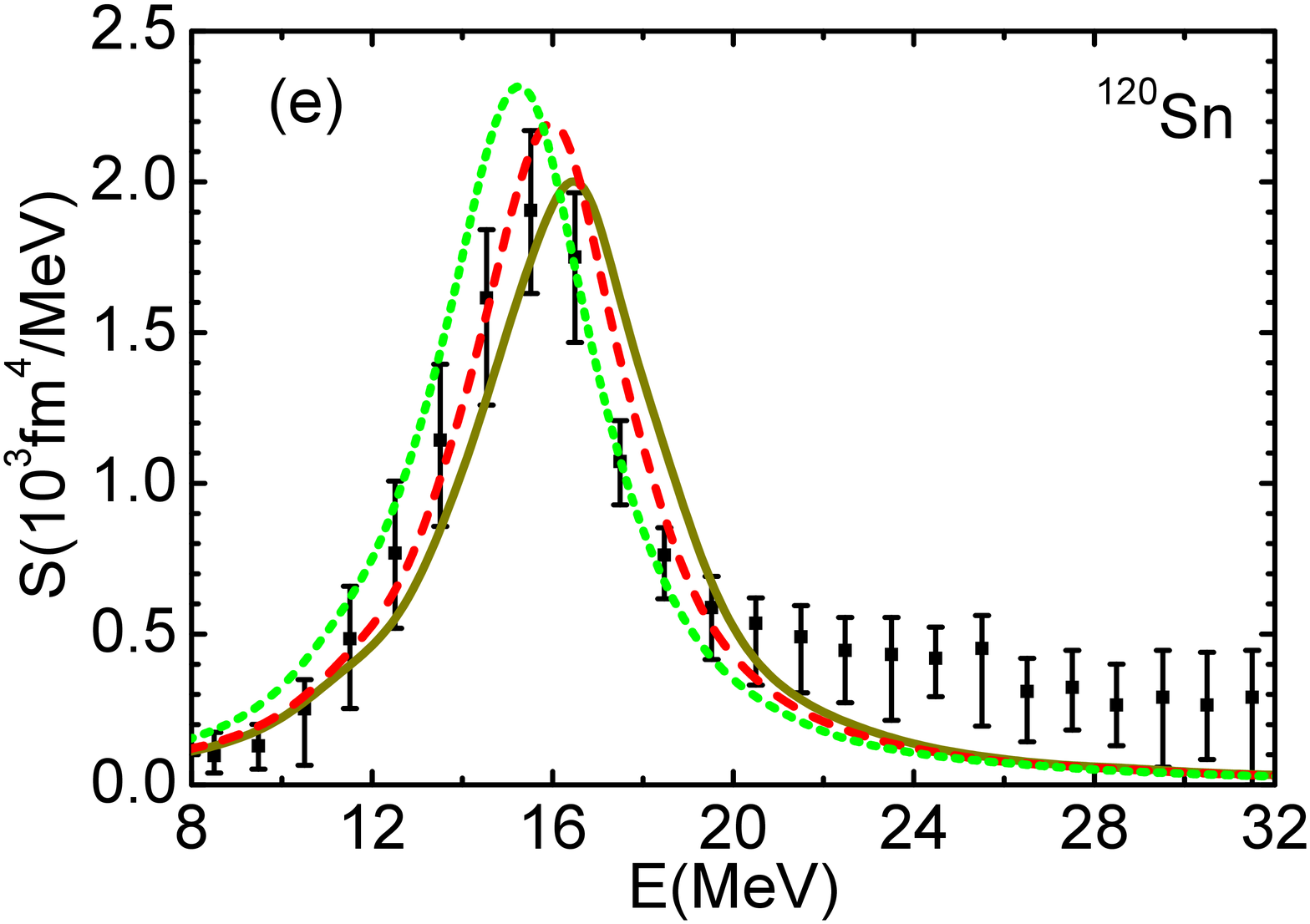}
\includegraphics[width=0.47\textwidth]{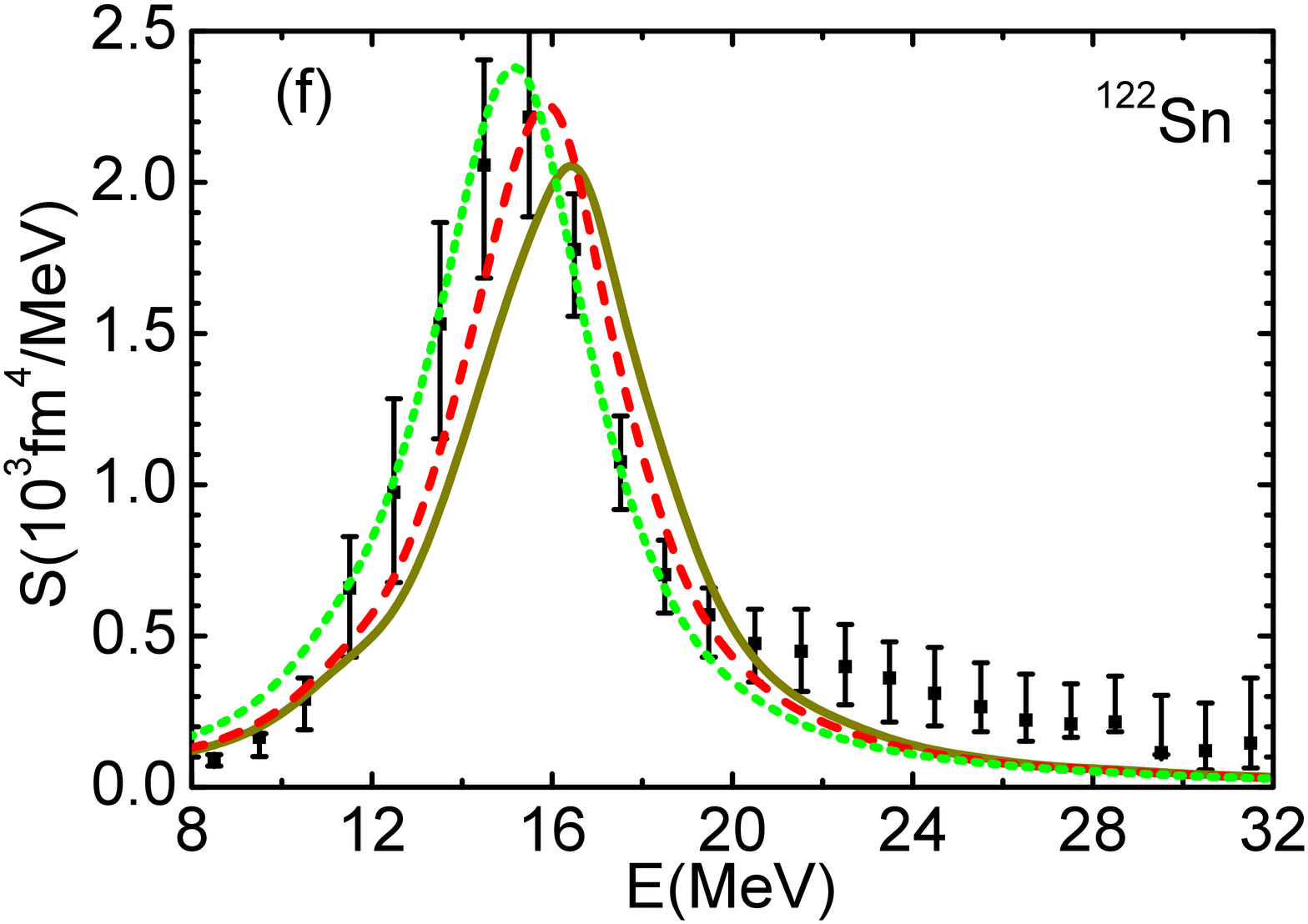}
\vglue -2cm
\includegraphics[width=0.47\textwidth]{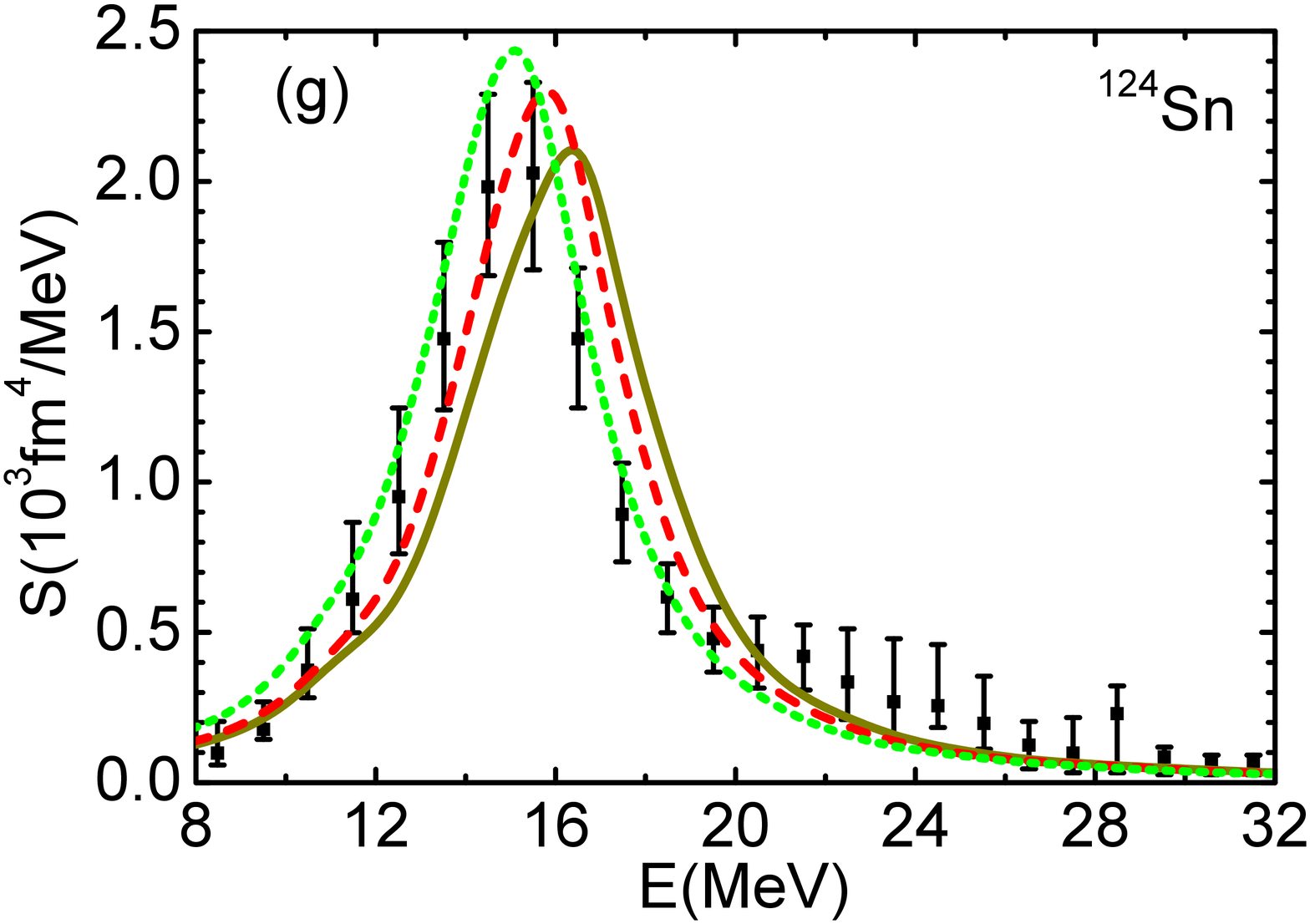}
\vglue -2cm
\caption{The calculated ISGMR strength distributions in
$^{112-124}$Sn are compared with the experimental data
from Refs. \cite{Li07,Li10}. The SLy5 (solid line), SkM*
(dashed line) and SKP (dotted line) forces are adopted
in the calculations whose results are shown here,
together with the mixed pairing interaction.}
\label{Fig.4}
\end{figure*}

\begin{figure*}[hbt]
\includegraphics[width=0.47\textwidth]{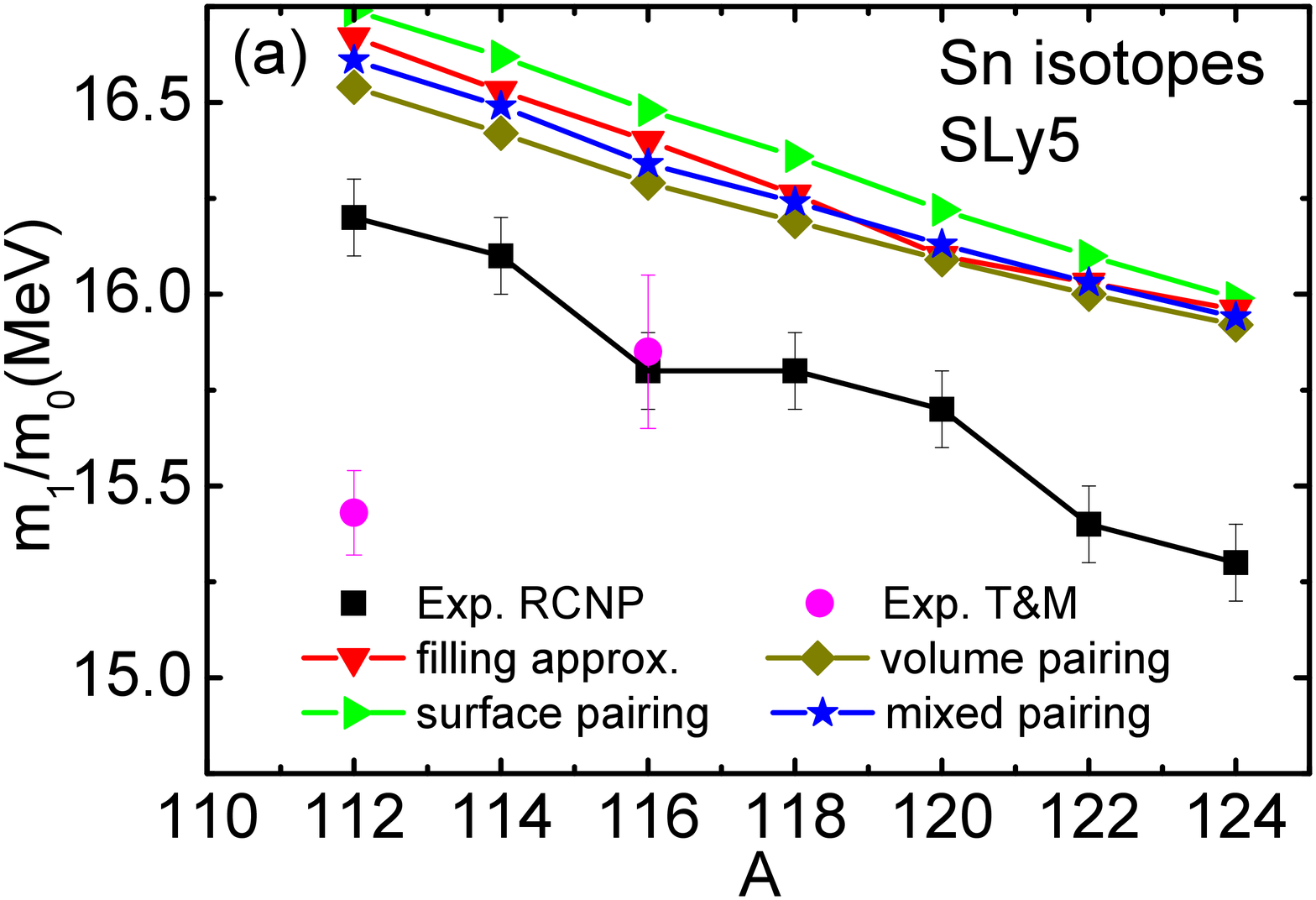}
\includegraphics[width=0.47\textwidth]{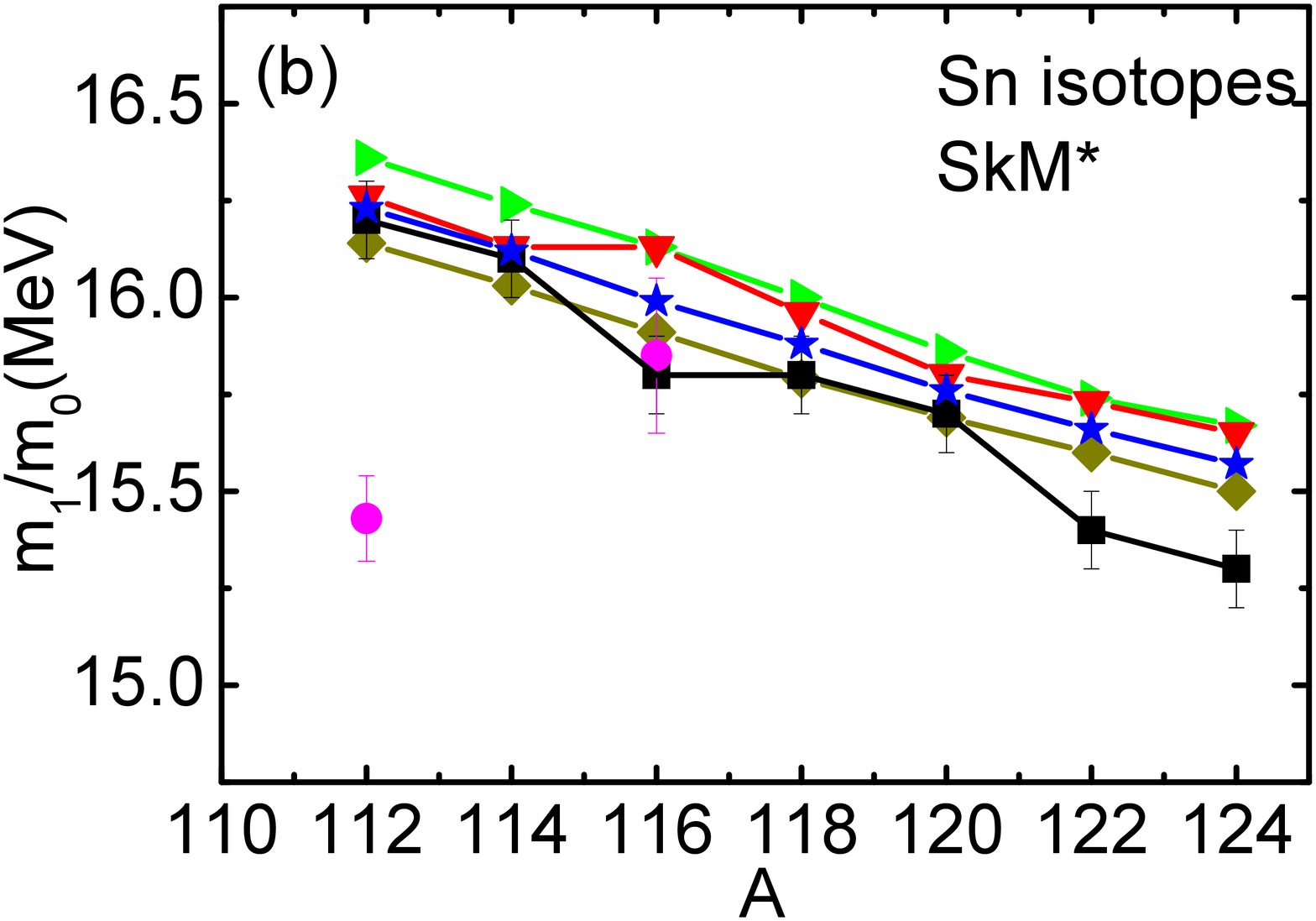}
\vglue -2cm
\includegraphics[width=0.47\textwidth]{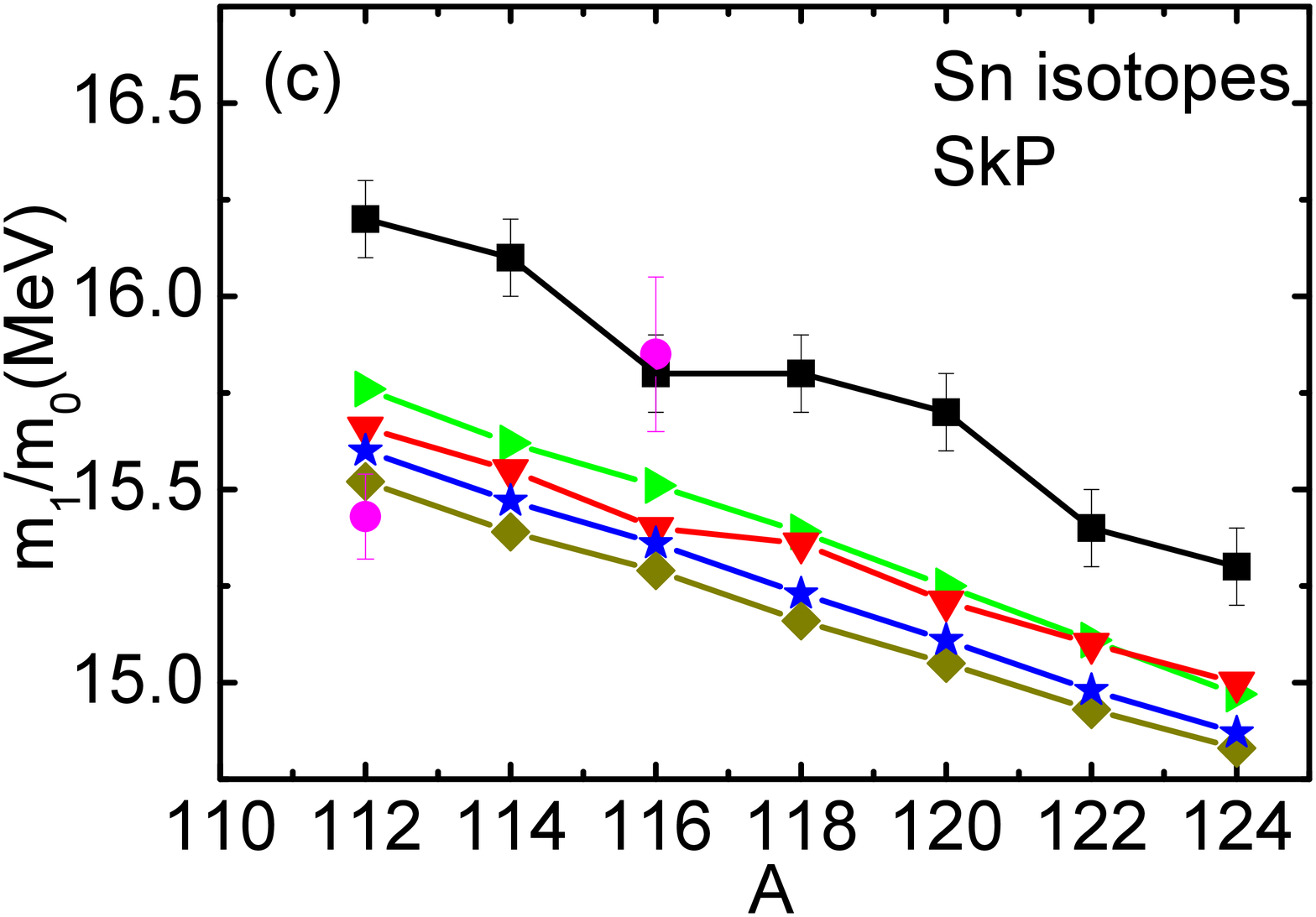}
\vglue -2cm
\caption{The calculated ISGMR centroid energies in the
even-even $^{112-124}$Sn isotopes are compared with
the experimental data from Refs. \cite{Li07,Li10}.
The forces SLy5 (a), SkM* (b) and SkP (c) are adopted
in the present calculations together with either the filling
approximation, or the volume, or the surface, or the mixed
pairing interactions, respectively.} \label{Fig.5}
\end{figure*}

\begin{table*}
\caption{The calculated ISGMR constrained energies
($E_{con}=\sqrt{m_1/m_{-1}}$), centroid energies ($E_{cen}=m_1/m_0$),
and scaling energies ($E_{s}=\sqrt{m_3/m_1}$) in even-even
$^{112-124}$Sn isotopes are compared with the experimental data.
The theoretical results are obtained in the interval between
10.5 and 20.5 MeV by using the SkP, SkM*, and SLy5 parameter sets
together with the mixed pairing interaction. The experimental data
are taken from Refs. \cite{Li07,Li10,Lui04,You04}. The
values in parenthesis are the difference between the
theoretical values and the experimental data. Units are
MeV.}
\begin{ruledtabular}
\begin{tabular}{ccccccccccccc}
                    &            &  Exp.    &   SkP  &  SkM*  &   SLy5     \\
\hline
                    & $^{112}$Sn & 16.1$\pm$0.1 & 15.55(-0.55) & 16.18(0.08) & 16.55(0.45)  \\
                    &            &15.23$^{+0.10}_{-0.10}$&  &   &        \\
                    & $^{114}$Sn & 15.9$\pm$0.1 & 15.42(-0.48) & 16.06(0.16) & 16.42(0.52)  \\
                    & $^{116}$Sn & 15.7$\pm$0.1 & 15.31(-0.39) & 15.94(0.24) & 16.29(0.59)   \\
$E_{con}=\sqrt{m_1/m_{-1}}$
                    & $^{118}$Sn & 15.6$\pm$0.1 & 15.17(-0.43) & 15.82(0.22) & 16.17(0.57)   \\
                    & $^{120}$Sn & 15.5$\pm$0.1 & 15.05(-0.45) & 15.71(0.21) & 16.05(0.55)   \\
                    & $^{122}$Sn & 15.2$\pm$0.1 & 14.92(-0.28) & 15.60(0.40) & 15.94(0.74)    \\
                    & $^{124}$Sn & 15.1$\pm$0.1 & 14.80 (-0.3)& 15.49(0.39) & 15.85(0.75)   \\
                    &            &14.33$^{+0.17}_{-0.14}$&  &   &          \\
\hline
                    & $^{112}$Sn & 16.2$\pm$0.1 & 15.60(-0.6) & 16.23(0.03) & 16.61(0.41)   \\
                    &           &15.43$^{+0.11}_{-0.10}$&   &    &       \\
                    & $^{114}$Sn & 16.1$\pm$0.1 & 15.27(-0.83) & 16.12(0.02) & 16.49(0.39)   \\
                    & $^{116}$Sn & 15.8$\pm$0.1 & 15.36(-0.44) & 16.00(0.20) & 16.36(0.56)   \\
                    &            &15.85$^{+0.20}_{-0.20}$&   &    &      \\
$E_{cen}=m_1/m_0$
                    & $^{118}$Sn & 15.8$\pm$0.1 & 15.23(-0.57) & 15.88(0.08) & 16.25(0.45)    \\
                    & $^{120}$Sn & 15.7$\pm$0.1 & 15.11(-0.59) & 15.78(0.08) & 16.13(0.43)  \\
                    & $^{122}$Sn &15.4$\pm$0.1 & 14.99(-0.41) & 15.67(0.27) & 16.03(0.63)   \\
                    & $^{124}$Sn &15.3$\pm$0.1 & 14.87(-0.43) & 15.57(0.27) & 15.95(0.65)   \\
                    &           &14.50$^{+0.14}_{-0.14}$&   &    &    \\
\hline
                    & $^{112}$Sn  & 16.7$\pm$0.2 & 15.74(-0.96) & 16.38(-0.32) & 16.77(0.07)   \\
                    &             &16.05$^{+0.26}_{-0.14}$ &   &    &       \\
                    & $^{114}$Sn & 16.5$\pm$0.2 & 15.62(-0.88) & 16.18(-0.32) & 16.66(0.16)    \\
                    & $^{116}$Sn &  16.3$\pm$0.2 & 15.53(-0.77) & 16.16(-0.14) & 16.54(0.24)   \\
$E_{s}=\sqrt{m_3/m_{1}}$
                    & $^{118}$Sn &  16.3$\pm$0.1 & 15.40(-0.9) & 16.05(-0.25) & 16.45(0.15)    \\
                    & $^{120}$Sn & 16.2$\pm$0.2 & 15.29(-0.91) & 15.96(-0.24) & 16.35(0.15)    \\
                    & $^{122}$Sn & 15.9$\pm$0.2 & 15.18(-0.71) & 15.86(-0.04) & 16.26(0.36)    \\
                    & $^{124}$Sn & 15.8$\pm$0.1 & 15.07(-0.73) & 15.76(-0.04) & 16.18(0.38)   \\
                    &          &14.96$^{+0.10}_{-0.11}$ &   &    &       \\
\end{tabular}
\end{ruledtabular}
\end{table*}

The mass number dependence of the calculated ISGMR centroid energies
in the Sn isotopes is shown in Fig. 5, for the case of
the SLy5, SkM* and SkP sets with the various pairing interactions.
The results obtained without pairing, by using the filling approximation,
are also shown in the same figure. It should be noticed that
in some cases the experimental data are not consistent with
each other. However, at the same time it is quite clear that
the predicted centroid energies obtained using the SLy5 and SkP
interactions are not in agreement with the experimental data:
the SLy5 results, even when the pairing effects are taken
into account, overestimate the experimental data while
the SkP results underestimate them. In the case of the SkM*
interaction, the predictions are much improved compared to
the case of SLy5 and SkP. The calculations with the volume and
mixed pairing forces (and the filling approximation) give
better predictions for $^{112,114}$Sn.
The results provided by the volume pairing force in
$^{116,118,120}$Sn are very close to the experimental data,
while they slightly overestimate the experimental findings in $^{122,124}$Sn.

There is some difference between the results presented here,
and those already discussed in Ref. \cite{Li08}. Compared to
 Ref. \cite{Li08}, we have included here the contribution of
the two-body spin-orbit interaction which provides an attractive
effect and lowers the RPA and QRPA ISGMR energies. However,
in order to be able to perform many systematic calculations,
we have chosen in the current work
the  QRPA on top of HF-BCS instead of
HFB. Pairing effects are slightly different in the two approaches,
so the discussion of this Section
on the results obtained with
either volume, or surface, or mixed pairing forces, is
not exactly the same as in Ref. \cite{Li08}. Despite these
differences, we have to stress that the overall qualitative
conclusion that the Sn data are rather consistent with the
value of nuclear incompressibility associated with SkM*
once the pairing is taken into account, is the same
as in Ref. \cite{Li08}.

This conclusion can be reinforced by the values reported
in Table III, that are the various calculated energies
obtained by means of the QRPA calculations performed
with the mixed pairing interaction. The interaction SkP
underestimates all these energies with respect to experiment,
whereas SLy5 overestimates them. The interaction SkM* gives
better results for both the constrained energies, the
centroid energies, and the scaling energies.

Thus, the discrepancy between the values of the nuclear
incompressibility extracted either from Sn or Pb data still
remains to some extent a puzzle; however, the pairing
effects need to be taken into account and reduce this discrepancy
to only $\approx$ 5\%.

\subsection{Pb Isotopes}

Measurements of the ISGMR strength distributions in the
even-even nuclei $^{204,206}$Pb have been done recently at
RCNP, Osaka University \cite{Fuji11}. Several measurements of
the ISGMR strength distribution in $^{208}$Pb had been performed
already in the past, e.g., at KVI (Netherlands) and
Texas A\&M University (USA). One of the motivations to perform
the new measurement on Pb isotopes at RCNP has been the study of
a conjecture, that is, the possible appearance in the ISGMR energies of
the so-called
mutually-enhanced-magicity (MEM) effect proposed by Lunney and
Zeldes in the context of the mass systematics \cite{Lun03,Zel83}.
According to this conjecture, the ISGMR energy in double closed
shell nuclei might be higher
if compared with the systematic values in its neighboring open shell
isotopes \cite{Khan091}.
Experimentally, the peak energies that have been obtained for the
$^{204-208}$Pb isotopes \cite{Fuji11} are
13.98, 13.94, and 13.90 MeV, respectively. From these experimental results,
we can not infer any kind of MEM effect in Pb isotopes.

\begin{figure*}[hbt]
\includegraphics[width=0.47\textwidth]{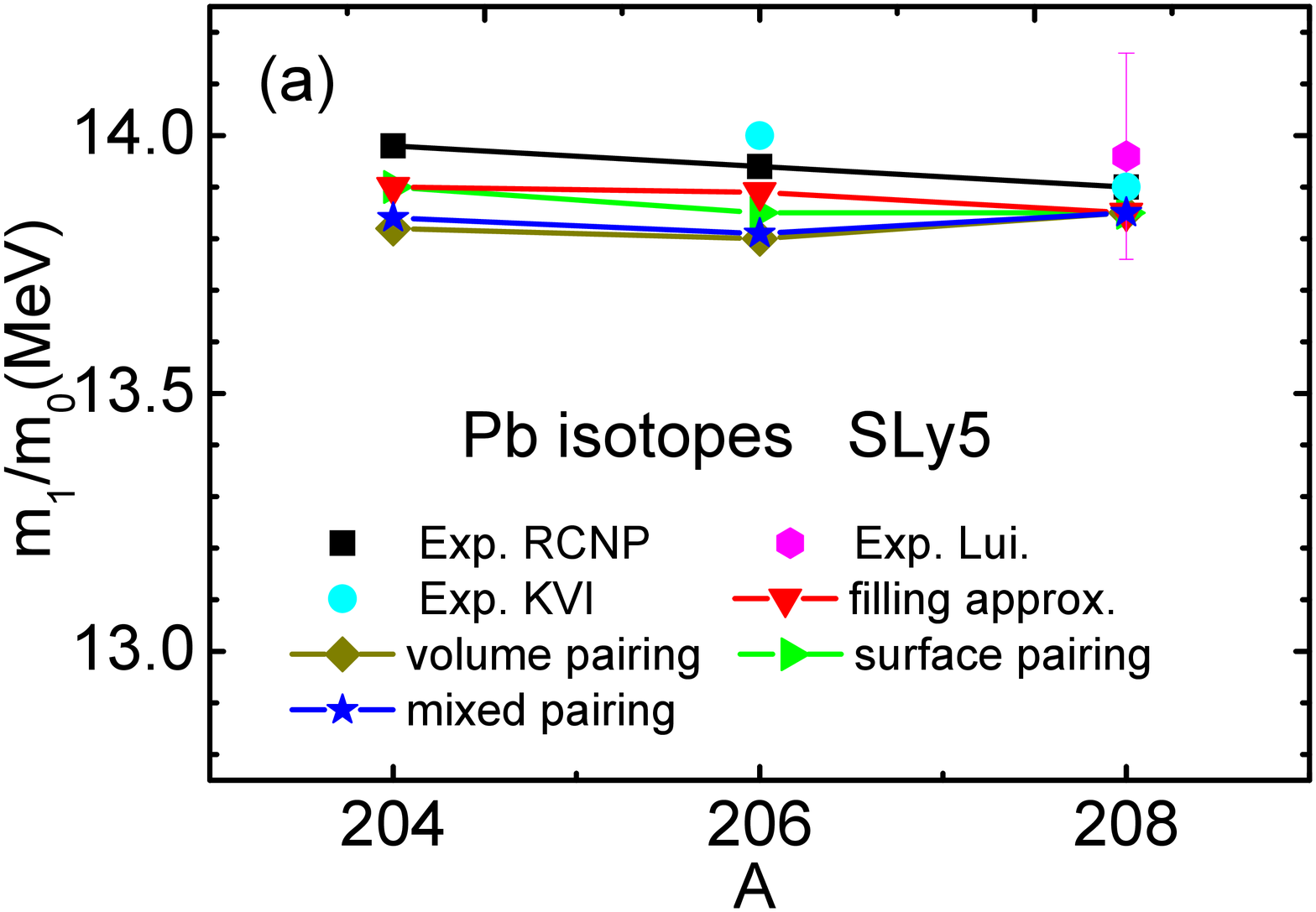}
\includegraphics[width=0.47\textwidth]{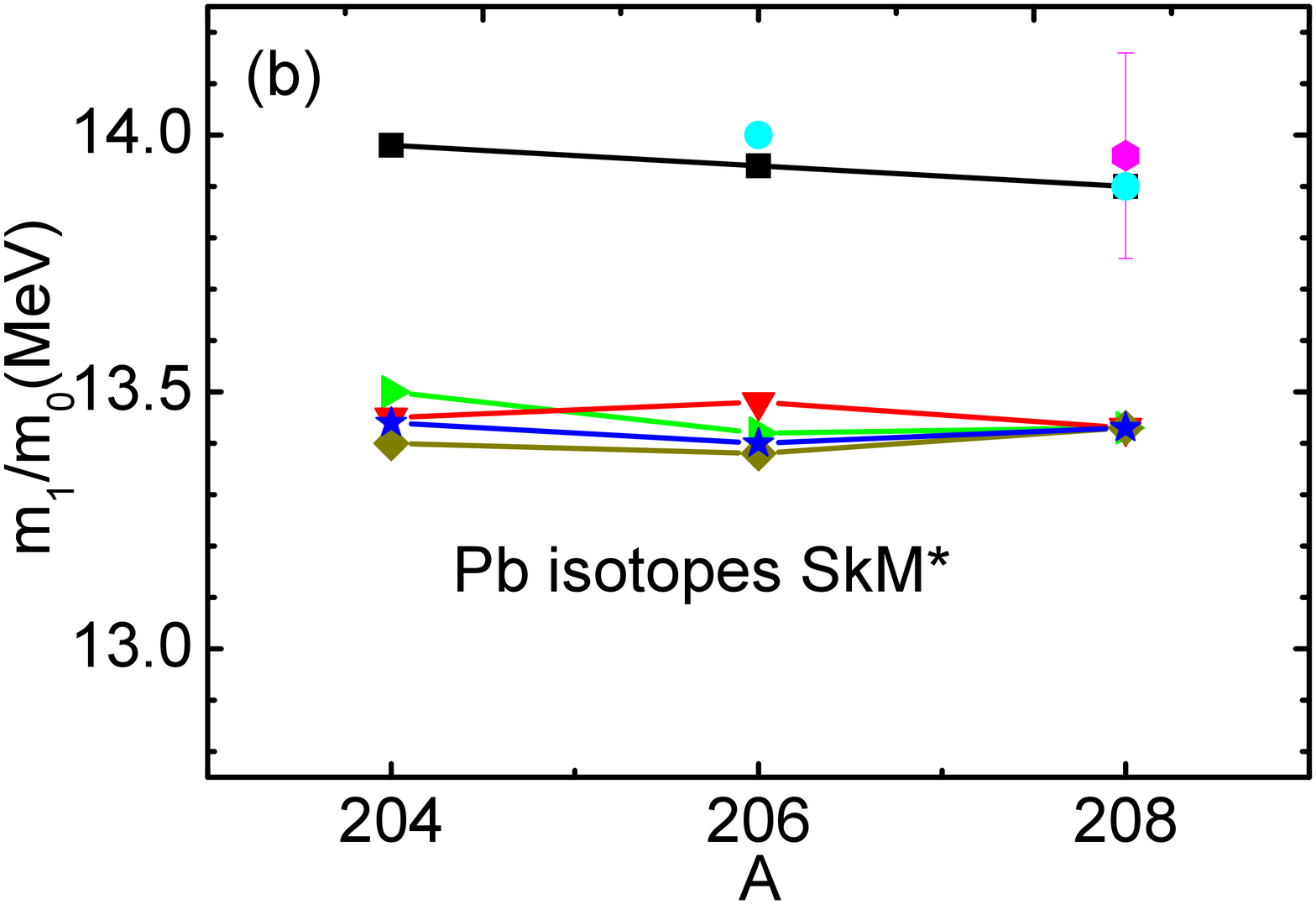}
\vglue -2cm
\includegraphics[width=0.47\textwidth]{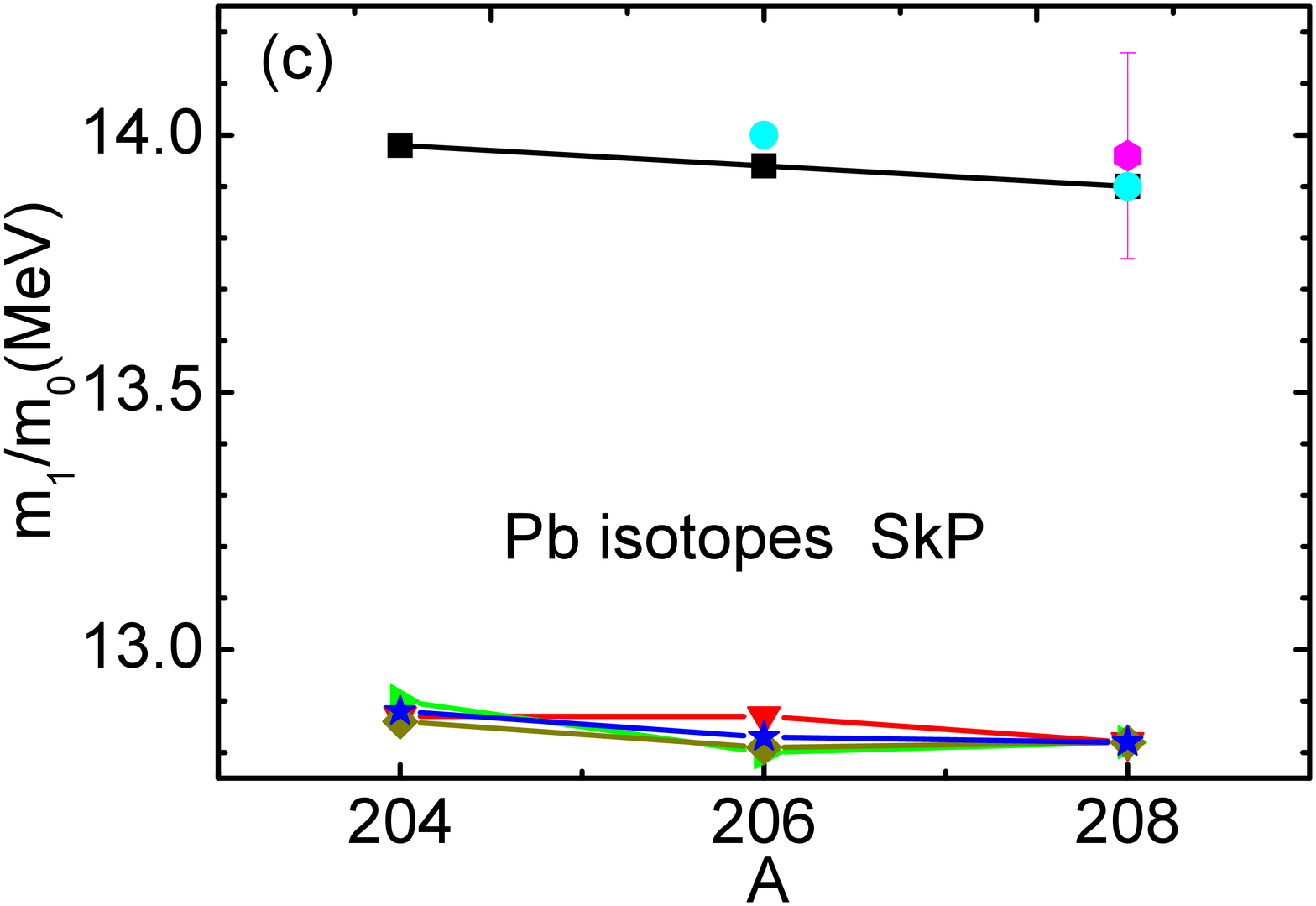}
\vglue -2cm
\caption{The calculated ISGMR centroid energies in the
even-even $^{204-208}$Pb isotopes are compared
with the experimental data obtained from Refs. \cite{Fuji11,Lui04,You04}.
The forces SLy5(a), SkM*(b) and SkP(c) are adopted in the present
calculations together with either the
filling approximation, or the volume, or the surface, or the
mixed pairing interactions, respectively.} \label{Fig.6}
\end{figure*}

\begin{table*}
\caption{The calculated ISGMR constrained energies
($E_{con}=\sqrt{m_1/m_{-1}}$), centroid energies ($E_{cen}=m_1/m_0$),
and scaling energies ($E_{s}=\sqrt{m_3/m_1}$) in even-even
$^{204,206,208}$Pb are compared with the experimental data.
The theoretical results are obtained in the interval between
10 and 20 MeV by using the SkP, SkM*, and SLy5 parameter sets,
together with the mixed pairing interaction. The experimental
data are taken from Refs. \cite{Fuji11,Lui04}.
The
values in parenthesis are the difference between the
theoretical values and the experimental data.
Units are MeV.}
\begin{ruledtabular}
\begin{tabular}{ccccccccccccc}
                    &     &  Exp.    &   SKP  &  SKM*  &   SLy5     \\

\hline
                    & $^{204}$Pb &               & 12.84 & 13.40 & 13.80   \\
$E_{con}=\sqrt{m_1/m_{-1}}$           & $^{206}$Pb &               & 12.79 & 13.36 & 13.76   \\
                    & $^{208}$Pb &               & 12.70 & 13.29 & 13.71   \\
\hline
                    & $^{204}$Pb &  13.98        & 12.88(-1.10) & 13.44(-0.54)  & 13.84(-0.14)   \\
$E_{cen}=m_1/m_0$   &$^{206}$Pb  &  13.94        & 12.83(-1.11) & 13.40(-0.54)  & 13.81(-0.13)    \\
                    &$^{208}$Pb  & 13.96$\pm$0.2 & 12.74(-1.22) & 13.34(-0.62)  & 13.77(-0.19)  \\
\hline
                    &$^{204}$Pb  &              & 13.01 & 13.56 & 13.98   \\
$E_{s}=\sqrt{m_3/m_1}$             &$^{206}$Pb  &              & 12.97 & 13.54 & 13.97    \\
                    &$^{208}$Pb  &              & 12.88 & 13.49 & 13.93   \\
\end{tabular}
\end{ruledtabular}
\end{table*}

In Fig. 6 and Table IV we show our theoretical results for the ISGMR
in Pb isotopes. The centroid energies calculated by using different
Skyrme sets and pairing forces are displayed in Fig. 6. We can see
that both SkM* and SkP underestimate the experimental data,
although the results obtained with SkM* are better.
The Skyrme force SLy5, having an associated value of
incompressibility K$_{\infty}$= 230 MeV, reproduce very well
the centroid energies in Pb isotopes. The same conclusion can be
drawn from the results shown in Table IV. It has been known
for some time that this value of K$_{\infty}$ can be extracted
from the $^{208}$Pb data, if the density dependence of the force is
the one that characterizes most of the recent Skyrme forces \cite{colo04}.
Indeed, from the present results
one can see that the effect of pairing in the Pb isotopes is rather
small. Finally, there is no evidence for any MEM effect in our
theoretical calculations which is  also confirmed by
the experimental results obtained at RCNP.

\section{Summary}

In summary, we have studied systematically the isoscalar
giant monopole resonance (ISGMR) in Cd, Sn and Pb isotopes
within the self-consistent Skyrme HF+BCS and quasi-particle
random phase approximation (QRPA).
Three different Skyrme parameter sets are used in the
present calculations, namely SLy5, SkM* and SkP: they are
chosen since they are characterized by different values of
the nuclear incompressibility, K$_{\infty}$=230, 217 and 202
MeV, respectively. To study the role of the pairing correlations,
we choose three types of pairing interactions,
i.e., the so-called volume, surface, and mixed pairing forces.
For the sake of comparison, we also produce RPA results (without
pairing) within the filling approximation.

The various kinds of centroid
energies, and the detailed strength distributions of the
ISGMR in Cd, Sn and Pb isotopes are compared with the
available experimental data. We have found that the pairing
correlations always decrease the peak
energies of the ISGMR because of the attractive
character of the particle-particle force in the 0$^+$ channel.
The typical size of this effect
is several hundreds keV.

From the present study, we find that the SkP interaction fails completely
in reproducing the ISGMR strength distributions in all the studied
isotopes due to its low value of nuclear matter incompressibility.
The SLy5 parameter set (having an associated incompressibility
of 230 MeV) gives a reasonable description
of the ISGMR in Pb isotopes, whereas a better overall description
in the case of Cd and Sn isotopes is achieved by using the
force SkM* (characterized by incompressibility of 217 MeV).
We have also found that the change  of the ISGMR energies  by
pairing correlations in $^{204,206}$Pb is quite small.
The results for the Pb isotopes suggest that both theoretically
and experimentally there is no  evidence for the so-called
MEM effect.

Pairing helps in reducing the discrepancy between the values
of the nuclear incompressibility extracted either from Pb or
Sn data. We also found that Cd data do not introduce further
problems as they seem to be rather consistent with Sn data.
A small discrepancy of about 5\% in K$_{\infty}$ between the conclusions drawn from Pb
and Sn data remains and may deserve further investigation.
Since the size of this discrepancy depends on the pairing
force and the model to treat its effects, one should better
analyze whether we can constrain the attractive particle-particle
matrix elements that appear in the QRPA calculations.
Pairing forces are usually constrained only by means of
ground-state properties, and at present it is unclear if
volume or surface pairing interactions should be preferred.
Along a different line, the present results may be
interpreted by saying that the role of surface and
surface-symmetry contributions to the nuclear incompressibility
are still not precisely fixed.

\section*{Acknowledgments}
We would like to thank M. Fujiwara, Y.-W.  Lui and U. Garg for
stimulating discussions and for providing us the experimental data
prior to publication. This work is partially supported by the Japanese
Ministry of Education, Culture, Sports, Science and Technology
by Grant-in-Aid for Scientific Research under
the program number (C(2))20540277. L.C. acknowledges the support
of the National Science Foundation of China under Grant Nos. 10875150
and 11175216. The support of the
Italian Research Project "Many-body theory of nuclear systems and
implications on the physics of neutron stars" (PRIN 2008) is
also acknowledged. This research was supported in part by the Project of Knowledge
Innovation Program (PKIP) of Chinese Academy of Sciences, Grant No. KJCX2.YW.W10.

\end{document}